\documentclass[twocolumn,secnumarabic,preprintnumbers,amsmath,amssymb,floatfix]{revtex4}
\bibpunct{[}{]}{;}{a}{,}{,}
\usepackage{graphicx}
\usepackage{dcolumn}
\usepackage{bm}
\usepackage[usenames]{color}

\begin{document}

\preprint{Special Issue of International Journal of Bifurcation and Chaos (Net-Work 2008)}

\title{Control of synchronization in coupled neural systems by time-delayed feedback}

\author{Philipp H\"{o}vel}
\author{Markus A. Dahlem}
\author{Eckehard Sch\"{o}ll}
\email{schoell@physik.tu-berlin.de}
\homepage{http://www.itp.tu-berlin.de/schoell}
\affiliation{Institut f{\"u}r Theoretische Physik, Technische Universit{\"a}t Berlin, Hardenbergstra{\ss}e 38, 10623
Berlin, Germany}

\date{\today}

\begin{abstract}
\noindent
We discuss the synchronization of coupled neurons which are modelled as FitzHugh-Nagumo systems. As smallest entity in a
larger network, we focus on two diffusively coupled subsystems, which can be interpreted as two mutually interacting
neural populations. Each system is prepared in the excitable regime and subject to independent random fluctuations. In
order to modify their cooperative dynamics, we apply a local external stimulus in form of an extended time-delayed
feedback loop that involves multiple delays weighted by a memory parameter and investigate if
local control applied to a subsystem can allow one to steer the global cooperative dynamics. Depending on the
choice of this new control parameter, we investigate different measures to quantify the influence on synchronization:
ratio of interspike intervals, power spectrum, interspike interval distribution, and phase
synchronization intervals. We show that the control method is more robust for increasing memory parameter.
\end{abstract}

\maketitle

\noindent 

\section{Introduction}
\label{sec:intro}
The network of neurons in the brain exhibits a subtle balance of dynamic chaos and selforganized order.
A number of neurological diseases like Parkinson or epilepsy are characterized by a disturbance of this balance, e.g.
synchronized firing of electrical pulses of the neurons \citep{SCH94e}. Modern concepts of time-delayed feedback control
have recently been applied to suppress this undesired synchrony \citep{ROS04,ROS04a,POP05,POP06,GAS07b,SCH07,GAS08}.

It was shown earlier \citep{HAU06} that application of time-delayed feedback, which was originally
suggested for
deterministic chaos control \citep{PYR92}, is able to influence the cooperative dynamics. As a measure of
cooperative dynamics, we consider coherence, timescales, and synchronization of noise-induced oscillations. With this
method, a control force is constructed from the difference of the current state of a system to its time-delayed
counterpart. Previously, it has also been used to influence noise-induced oscillations of a single excitable system
\citep{JAN03,BAL04,PRA07,POT08}, of systems below a Hopf bifurcation \citep{SCH04b,POM05a,POT07,FLU07}
or a global bifurcation
\citep{HIZ06,HIZ08}, and of spatially extended reaction-diffusion systems \citep{STE05a,BAL06,SCH08a}.

In this paper, we extend the work of \citet{HAU06} by application of a different
feedback stimulation with multiple time delays. External delayed feedback loops have
been suggested for suppression of pathological brain rhythms \citep{ROS04,POP06}. Our method, also known as extended
time-delayed feedback, was initially proposed by Socolar et al. in order to extend the domain of effective
stabilization of unstable periodic orbits \citep{SOC94}. It generalizes the Pyragas scheme by introducing an additional
memory parameter and is known for successful stabilization at a larger range of parameters compared to the Pyragas
method \citep{UNK03,SCH03a,DAH07,DAH08b}. In case of noise-induced oscillations, extended feedback has
been demonstrated to result in drastically improved coherence and arbitrarily long correlation times \citep{POM07}. 

Our aim is to control the global cooperative dynamics of the ensemble of neural populations by local application of a
stimulus to a single subsystem that involves multiple delays in the feedback.
In particular, we are interested in the study of the effects of the memory parameter as new control
parameter on the synchrony properties in coupled neural systems.
We compare this to the case of vanishing memory parameter, as investigated by \citet{HAU06}, and find that the control
scheme is more robust for multiple time-delayed feedback in the sense that enhanced synchronization is independent on
the tuning of the time delay.

The structure of the paper is the following: In Sec.~\ref{sec:model}, we introduce the model equation and the control
method. In Sec.~\ref{sec:ISI}, we discuss the configuration of the uncontrolled system and consider the average
interspike intervals. Sections~\ref{sec:spectrum} and \ref{sec:distr} are
devoted to the effects of the control parameters on the power spectrum and the interspike interval
distribution, respectively. In Sec.~\ref{sec:APS}, we introduce a phase variable and discuss effects of phase
synchronization. Finally, we
conclude with Sec.~\ref{sec:conclusion}.

\section{Model}
\label{sec:model}
In the following, we consider two mutually coupled neurons modelled by FitzHugh-Nagumo systems:
\begin{subequations}
\label{eq:FHN_ETDAS}
\begin{eqnarray}
	\label{eq:FHN_ETDAS_1}
	\epsilon_1 \dot{x}_1(t) &=& x_1(t) -\frac{x_1^3(t)}{3} -y_1(t) + C [x_2(t) -x_1(t)] \nonumber\\
	\dot{y}_1(t) &=& x_1(t) + a + D_1 \xi_1(t) \\
	&\,& + K \sum_{n=0}^{\infty} R^n \left[y_1(t-(n+1)\tau) - y_1(t-n \tau) \right]  \nonumber\\
	\label{eq:FHN_ETDAS_2}
	\epsilon_2 \dot{x}_2(t) &=& x_2(t) -\frac{x_2^3(t)}{3} -y_2(t) + C [x_1(t) -x_2(t)]\nonumber\\
	\dot{y}_2(t) &=& x_2(t) + a + D_2 \xi_2(t), 
\end{eqnarray}
\end{subequations}
where $x_1, y_1$ and $x_2, y_2$ correspond to single excitable systems representing neurons or
neural populations, which are diffusively coupled in the activator variables $x_1$,
$x_2$ with coupling strength $C$. The variables $y_1$ and $y_2$ represent the
inhibitor. Throughout this paper, we consider neurons in the excitable regime at which no autonomous oscillations
occur. Thus, we fix the excitability parameter $a$ as $a=1.05$.

In order to introduce different timescales in the two subsystems, we choose $\epsilon_1=0.005$ and $\epsilon_2=0.1$.
Each neuron is driven by Gaussian white noise $\xi_i(t) (i=1,2)$ with zero mean and unity variance. The noise
intensities are denoted by parameters $D_1$ and $D_2$, respectively, where we keep $D_2$ fixed at $D_2=0.09$ in the
following. The last term in Eq.~(\ref{eq:FHN_ETDAS_1}) describes extended time-delayed feedback control \citep{SOC94}
with time delay $\tau$, feedback gain K, and memory parameter $R$. Note that there exists an equivalent
recursive form of the feedback $F(t)$:
\begin{eqnarray}
 F(t) &=& K \sum_{n=0}^{\infty} R^n \left[y_1(t-(n+1)\tau) - y_1(t-n \tau) \right] \\
 &=& K \left[y_1(t-\tau) - y_1(t) \right]+RF(t-\tau). 
\end{eqnarray} 
The latter form is more amenable to experimental realization and for practical 
applications because the delayed feedback force $F(t-\tau)$ replaces the infinite series.

\subsection{Moderate, weak, and strong synchronization}
\label{subsec:mws} 
The two neurons or neural populations of Eqs.~(\ref{eq:FHN_ETDAS}) are prepared in the excitable regime. Without
external input, they remain in their stable fixed points. Random fluctuations lead to spiking. 

\begin{figure}[th]
  \begin{center}
	\includegraphics[width=\linewidth]{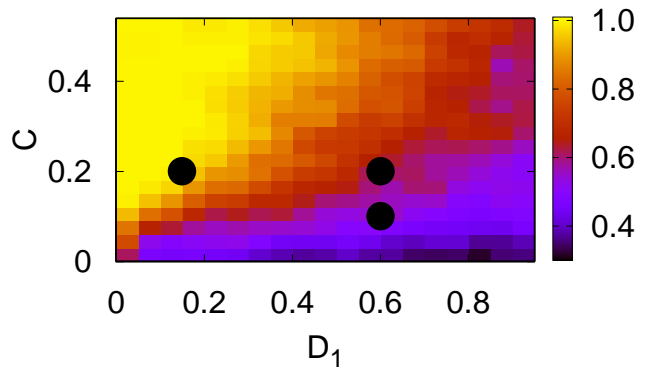}
  \end{center}
  \caption{\label{fig:isi_D1_C}(Color online) Ratio of interspike intervals $\langle T_1\rangle /\langle T_2\rangle $ of
the two subsystems in dependence on the coupling strength $C$ and noise intensity $D_1$. No control is applied to the
system. The dots mark the parameter choice for different synchronization regimes used in the following. Other
parameters as in Fig.~\ref{fig:x_t_mod_weak_strong}.}
\end{figure}

For reason of comparison, we consider at first the case where no control is applied to the system. We set the noise
intensity $D_2$ in the second subsystem to a small value, $D_2=0.09$, to realize some background noise
level. Depending on the coupling strength $C$ and the noise intensity $D_1$, the two subunits show cooperative dynamics.

Calculating the average interspike interval of the two neurons $\langle T_1\rangle$ and $\langle T_2\rangle $ and their
ratio, one can see in Fig.~\ref{fig:isi_D1_C} how the frequency synchronization changes in dependence on the coupling
strength and noise intensity. For a small value of $D_1$ and large coupling strength, the two subsystems display
synchronized behavior, $\langle T_1\rangle/\langle T_2\rangle\approx 1 $. On average, they show the same number of
spikes indicated by bright (yellow) color.

Since we are interested in the effects of a control force on the synchronization, in the following, we consider three
different cases: moderately, weakly, and strongly synchronized systems given by the specific choices of the coupling
strength and noise intensity in the first subsystem ($C=0.2$, $D_1=0.6$),  ($C=0.1$, $D_1=0.6$), and ($C=0.2$,
$D_1=0.15$), respectively. These different cases of stochastic synchronization are marked as black dots in
Fig.~\ref{fig:isi_D1_C}. Strong synchronization can be found for small noise intensity $D_1$ and large $C$, e.g.,
$D_1=0.15$ and $C=0.2$. Moderate synchronization is given for a choice of $D_1=0.6$ and $C=0.2$, and weak
synchronization can be realized by $D_1=0.6$ and $C=0.1$.

\begin{figure}[th]
  \begin{center}
	\includegraphics[width=\linewidth]{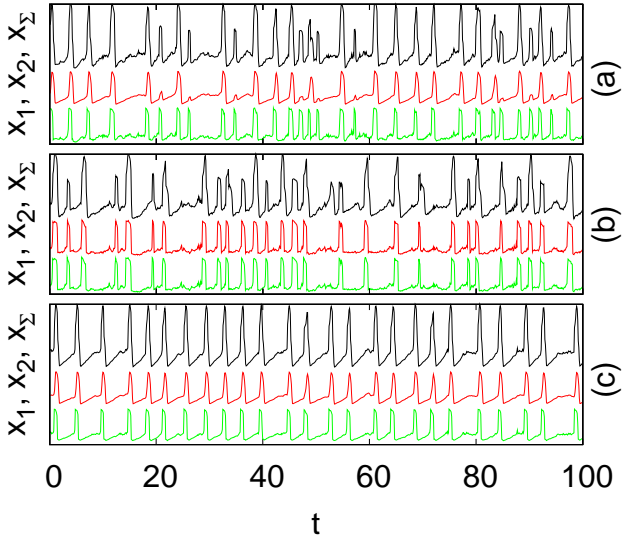}
  \end{center}
  \caption{\label{fig:x_t_mod_weak_strong}(Color online) Time series of the coupled FitzHugh-Nagumo system in the
absence of control. Panels (a), (b), and (c) correspond to moderately ($C=0.2$, $D_1=0.6$), weakly ($C=0.1$, $D_1=0.6$),
and strongly ($C=0.2$, $D_1=0.15$) synchronized systems, respectively. In all panels, the black, grey (red), and
lightgrey (green) curves refer to the summarized variable $x_\Sigma=x_1+x_2$, the $x_2$-, and the $x_1$-variable,
respectively. Other parameters: $a=1.05$, $\epsilon_1=0.005$, $\epsilon_2=0.1$, and $D_2=0.09$.}
\end{figure}

In terms of the time series, the three configurations of moderate, weak, and strong synchronization are displayed in
Fig.~\ref{fig:x_t_mod_weak_strong} as panels (a), (b), and (c), respectively. In all panels, the black, grey (red), and
lightgrey (green) curves refer to the global $x_\Sigma$-variable, the $x_2$-, and $x_1$-variable, respectively, where
$x_\Sigma$ is given by $x_\Sigma=x_1+x_2$. 

Moderately synchronized systems perform mostly synchronized spiking. However, there are certain events where only one
system shows an oscillation (see panel (a) of Fig.~\ref{fig:x_t_mod_weak_strong}). In the case of weak synchronization,
the spikes of the two subsystems coincide less as can be seen from the time series of the summarized signal $x_\Sigma$
in panel (b) of Fig.~\ref{fig:x_t_mod_weak_strong}. For strongly synchronized subsystems (see panel (c) of
Fig.~\ref{fig:x_t_mod_weak_strong}), the time series of the $x_1$- and $x_2$-variable exhibit spiking at the same time. 

\subsection{Time series and control}
\label{subsec:time_control}
 
\begin{figure}[th]
  \begin{center}
	\includegraphics[width=\linewidth]{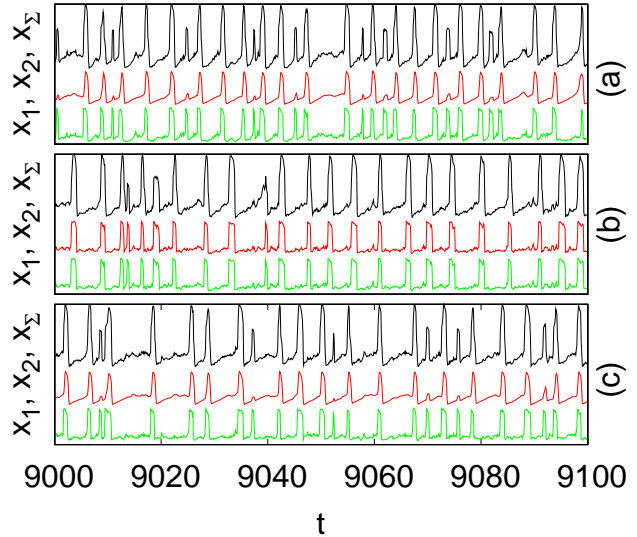}
  \end{center}
  \caption{\label{fig:x_t_mod_R}(Color online) Time series of the coupled FitzHugh-Nagumo system for moderate
synchronization. Panel (a) corresponds to no control. In panels (b) and (c), time-delayed feedback is applied
to the system with different memory parameters $R=0$ and $R=0.9$, respectively. Other control parameters are fixed at
$\tau=1$ and $K=1.5$. Grayscale (color coding) and other parameters as in Fig.~\ref{fig:x_t_mod_weak_strong}.}
\end{figure}

Equation~(\ref{eq:FHN_ETDAS_1}) already includes the time-delayed feedback scheme, whose parameters are the feedback
gain $K$, the time delay $\tau$, and the memory parameter $R$. The special case of $R=0$, also known as Pyragas control
\citep{PYR92}, was already investigated by \citep{HAU06}. Therefore, in the present work, we
discuss an extension to a different feedback stimulation involving multiple delays. In this extended
form, the method generates a control force from the differences of the states of the system that are one time unit
$\tau$ apart. The memory parameter $R \in (-1,1)$ can be understood as a weight of states that are further in the past.
For vanishing memory parameter $R=0$, only one delayed state enters the generation of the control force. 

We consider the case where the control force is applied to the first system only and in the inhibitor variable. Note
that one could also realize a feedback scheme that is applied to both subsystems and study effects of different values
of the control parameters for each subsystem, but these investigations are out of the scope of this work. Here we put
special emphasis on the effects due to changes in the memory parameter.

Figure~\ref{fig:x_t_mod_R} depicts different time series for moderate synchronization (C$=0.2$, $D_1=0.6$) in the
absence of control (panel(a)) compared to the cases when time-delayed is applied to the system (panels (b) and (c) for
memory parameters of $R=0$ and $R=0.9$). The time delay and feedback gain are chosen as $\tau=1$ and $K=1.5$. In all
panels, the black, grey (red), and lightgrey (green) curves correspond to the summarized global signal
$x_\Sigma$, the $x_2$-, and the $x_1$-variable, respectively. One can see that the time-delayed feedback enhances the
synchronization, i.e., events at which only one system oscillates are less frequent. In this sense the choice of $R=0$
is more efficient compared to larger memory parameters.

\section{Ratio of Average Interspike intervals}
\label{sec:ISI}
As first measure to quantify changes in the synchronization due to the control force, in this section, we consider the
ratio of average interspike intervals. In the presence of a control force, i.e., $K\neq 0$, the cooperativity can be
influenced by the control parameters. Varying the feedback gain $K$, the time delay $\tau$, and the memory parameter
$R$, the average interspike interval can be altered. The case of $R=0$ was discussed in Ref.~\citep{HAU06}. 

\begin{figure*}[th]
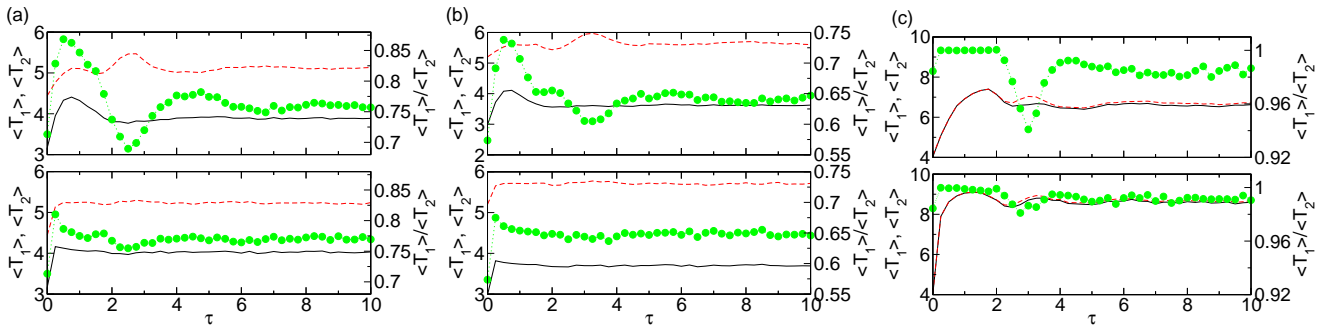

  \begin{center}
	\includegraphics[width=0.32\linewidth]{figure4a.eps}
	\includegraphics[width=0.32\linewidth]{figure4b.eps}
	\includegraphics[width=0.32\linewidth]{figure4c.eps}
  \end{center}
  \caption{\label{fig:isi_tau}(Color online) Interspike intervals in dependence on the time delay $\tau$. The green
dots correspond to the ratio of the interspike intervals $\langle T_1\rangle /\langle T_2\rangle $ of the two
subsystems, which are also depicted by solid (black) and dashed (red) curves for $\langle T_1\rangle$ and $\langle
T_2\rangle $, respectively. Panels (a), (b), and (c) correspond to the case of moderate, weak, and strong
synchronization, respectively. The control parameters are chosen as $K=1.5$ and the memory parameter corresponds to
$R=0$ and $0.9$ in the top and bottom panels, respectively. Other parameters as in Fig.~\ref{fig:x_t_mod_weak_strong}.}
\end{figure*}

For fixed feedback gain $K=1.5$ and two different values of $R$ ($R=0$ and $R=0.9$), Fig.~\ref{fig:isi_tau} depicts the
average interspike intervals of the subsystems, shown as solid (black) and dashed (red) curves for $\langle T_1\rangle$
and $\langle T_2\rangle $, and their ratio (as dotted (green) curve) for the case of moderately, weakly, and strongly
synchronized systems, respectively, in dependence on the time delay $\tau$. The left, middle, and right panels
correspond to the case of moderate, weak, and strong synchronization, respectively.

In all cases, tuning of the time delay leads to either enhanced and deteriorated synchronization for $R=0$. If
more information of past states ($R\neq 0$) is included, however, the variation of the ratio $\langle T_1\rangle
/\langle T_2\rangle $ is less sensitive to the specific choice of $\tau$. The bottom panels of Fig.~\ref{fig:isi_tau} do
not show large deviations for the ratio of average interspike intervals. Therefore, a larger memory parameter renders
the control method more robust.

\begin{figure*}[th]
  \begin{center}
	\includegraphics[width=0.32\linewidth]{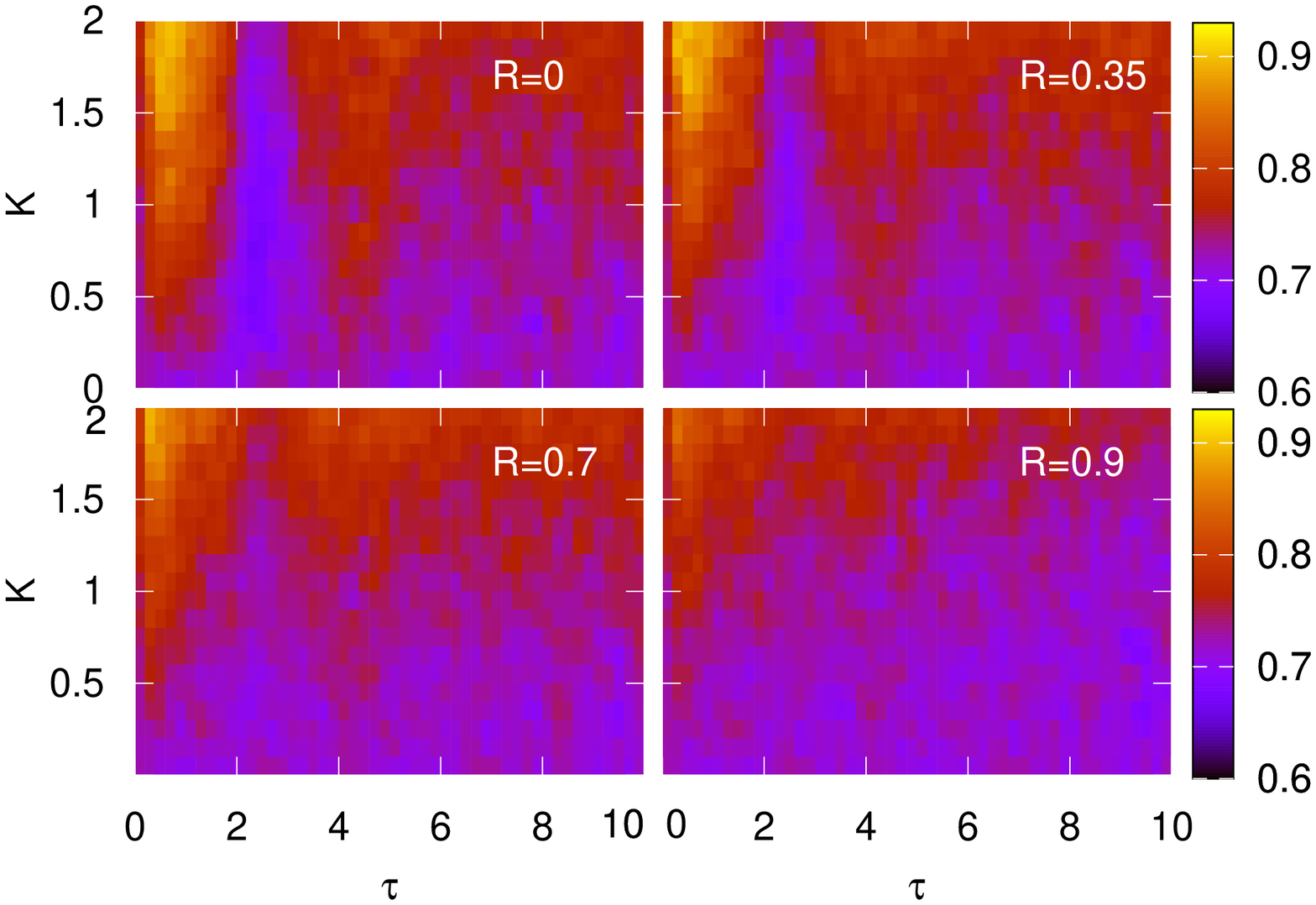}
	\includegraphics[width=0.32\linewidth]{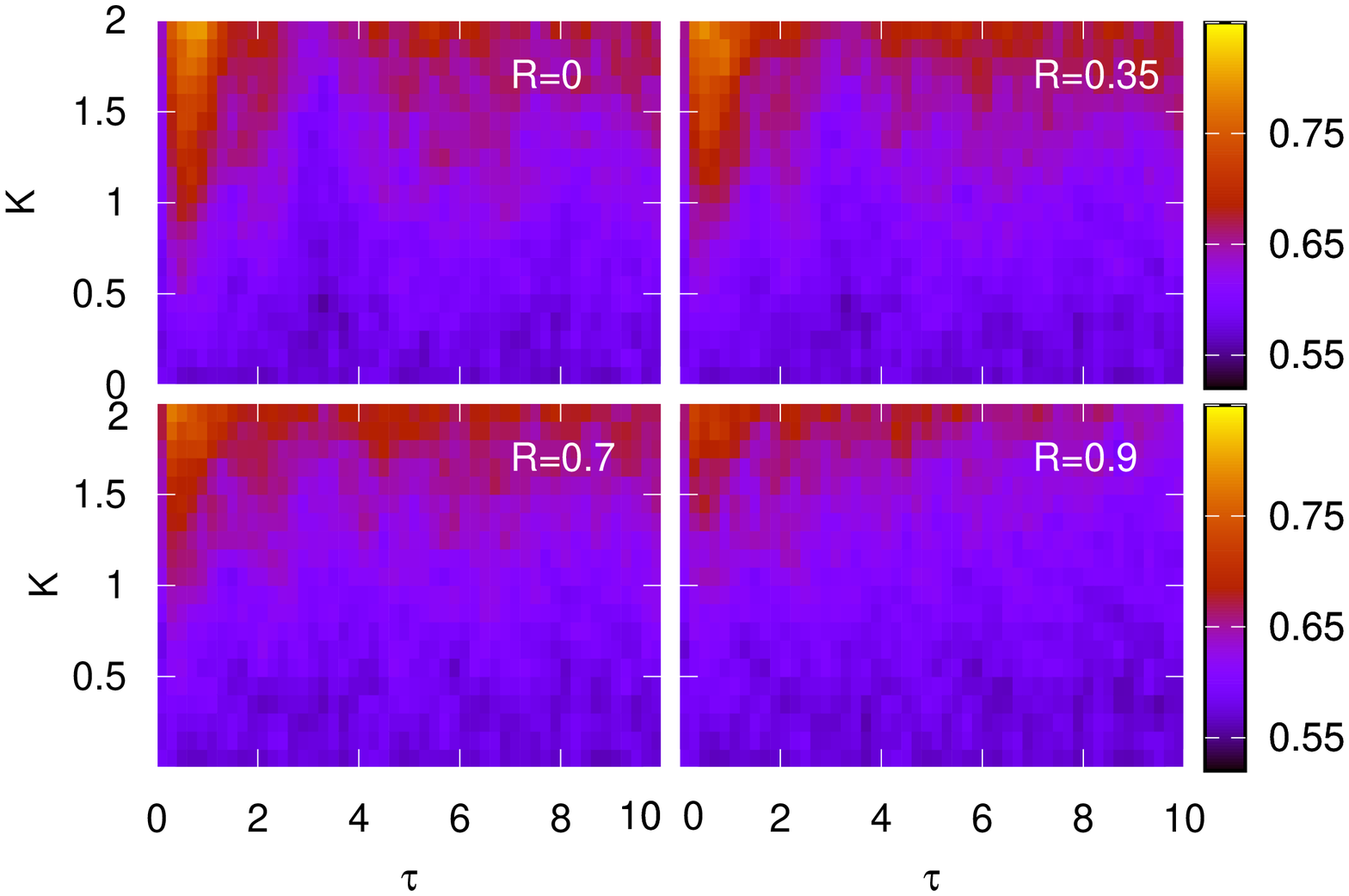}
	\includegraphics[width=0.32\linewidth]{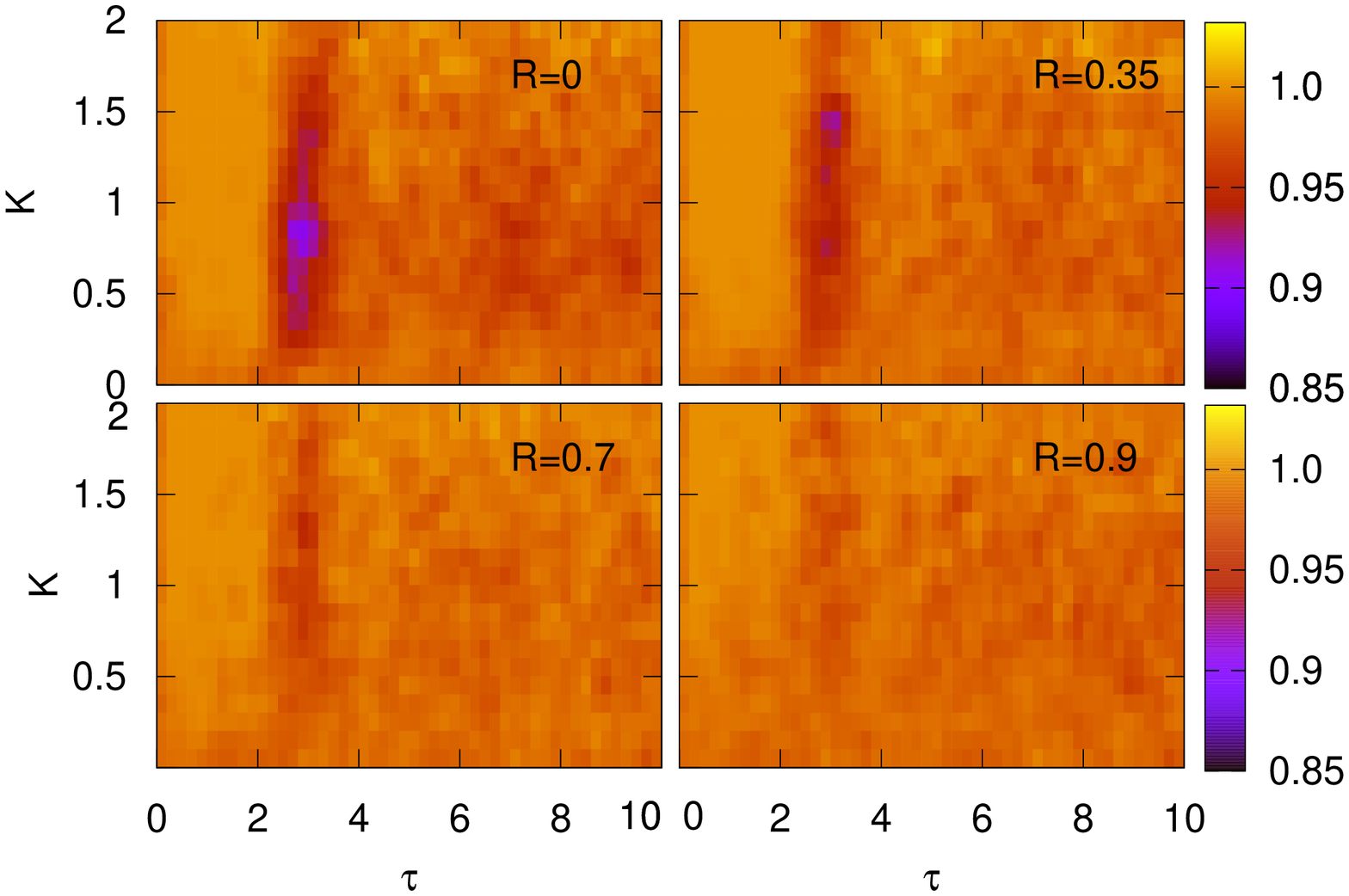}
  \end{center}
  \caption{\label{fig:isi_control}(Color online) Ratio of interspike intervals $\langle T_1\rangle / \langle
T_2\rangle$ in dependence on the feedback gain $K$ and the time delay $\tau$ for moderate ($C=0.2$, $D_1=0.6$), weak
($C=0.1$, $D_1=0.6$), and strong ($C=0.2$, $D_1=0.15$) synchronization in the left, middle, and right panel,
respectively. The memory parameter are fixed ar $R=0$, $0.35$, $0.7$, and $0.9$ in the four subfigures of all three
panels. Other parameters as in Fig.~\ref{fig:x_t_mod_weak_strong}.}
\end{figure*}

One can also consider two-dimensional projections of the control parameter space spanned by $K$, $\tau$, and $R$.
Parameterized by the feedback gain and time delay, Fig.~\ref{fig:isi_control} displays the ratio of $\langle T_1\rangle$
and $\langle T_2\rangle $ for moderate, weak, strong synchronization, respectively. The four panels in each figure
correspond to a memory parameter of $R=0$, $0.35$, $0.7$, and $0.9$. Note that Fig.~\ref{fig:isi_tau} can be understood
as horizontal cuts for $K=1.5$ in the respective diagram of Fig.~\ref{fig:isi_control}.

As also shown in the one-dimensional projections (see  Fig.~\ref{fig:isi_tau}), an increase of the memory parameter $R$
at fixed $K$ yields smaller changes of the ratio $\langle T_1\rangle / \langle T_2\rangle $. Independent on the feedback
gain $K$, the desynchronized darker region in Fig.~\ref{fig:isi_control} at a time delay $\tau \approx 3$ is much less
pronounced. For $R=0.9$, the ratio of average interspike intervals is constant in a wide range of the whole control
domain, which reflects the robustness of this extended feedback method.

\section{Power spectrum}
\label{sec:spectrum}
The two uncontrolled neural populations have different timescales, i.e., $\epsilon_1\neq\epsilon_2$, and
therefore, oscillate at different frequencies. We have shown in the last section that the mean frequencies of the
subsystems measured by the average interspike intervals match for certain control parameters. In order to gain a better
understanding of the shift in the timescales in the subsystem in the presence of control, we investigate the power
spectrum in this section, where we focus on the role of the memory parameter. 

\begin{figure*}[th]
  \begin{center}
	\includegraphics[width=\linewidth]{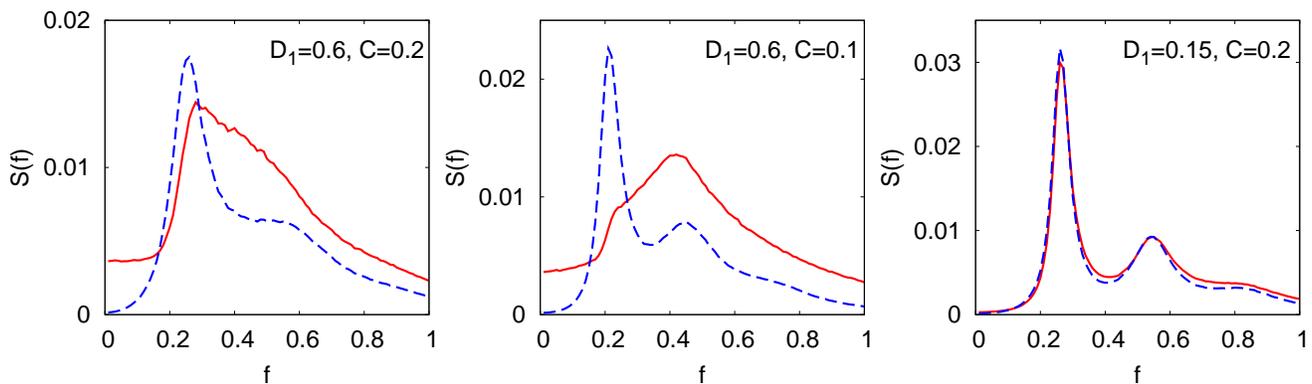}
  \end{center}
  \caption{\label{fig:spectrum_nocontrol}(Color online) Power spectrum of the two subsystems in the absence of control.
The left, middle, and right panels correspond to the case of moderate, weak, and strong synchronization,
respectively. The solid (red) and dashed (blue) curves refer to the $x_1$- and $x_2$-variable, respectively. Other
parameters as in Fig.~\ref{fig:x_t_mod_weak_strong}.}
\end{figure*}

At first, we consider the case when no control is applied to the system, i.e., $K=0$. The result can be seen in
Fig.~\ref{fig:spectrum_nocontrol} where the solid (red) and dashed (blue) curves correspond to the power spectrum, of
the $x_1$- and $x_2$-variable, respectively. The left, middle, and right panel refer to the case of moderate, weak,
and strong synchronization, respectively. This change of frequency synchronization is reached by choosing different
coupling strengths $C$ and noise intensities $D_1$. In the case of moderate synchronization ($C=0.2$, $D_1=0.6$) the
peaks of the power spectrum partly overlap. Weak synchronization ($C=0.1$, $D_1=0.6$) shows a smaller overlap, whereas
for strong synchronization ($C=0.2$, $D_1=0.15$) the spectra almost coincide.

\begin{figure*}[th]
  \begin{center}
	\includegraphics[width=\linewidth]{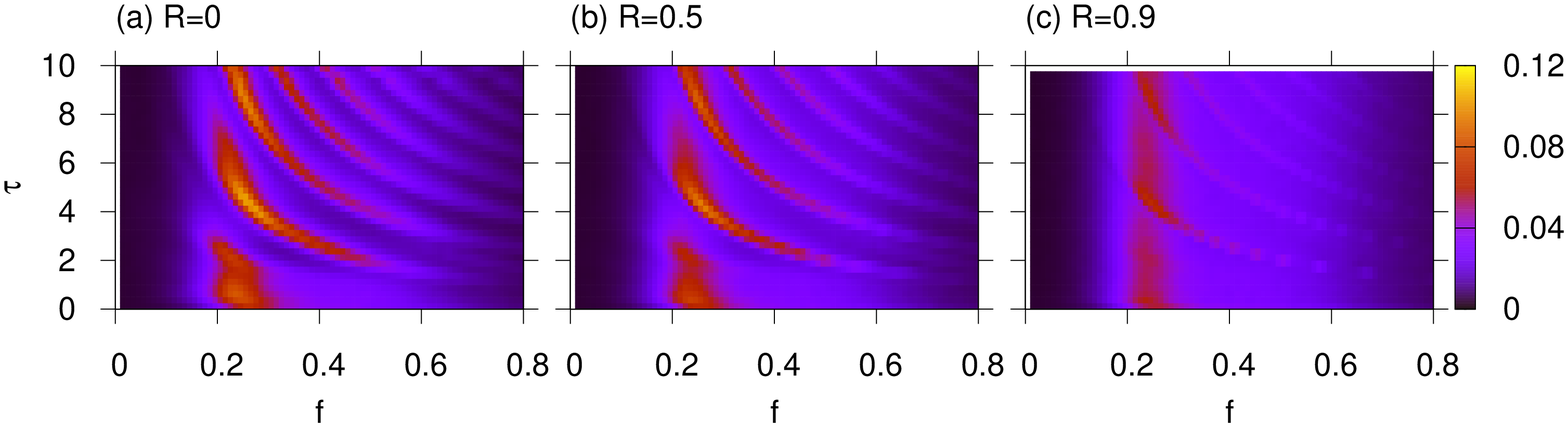}
  \end{center}
  \caption{\label{fig:spectrum_mod}(Color online) Power spectrum of the summarized signal $x_{\Sigma}=x_1+x_2$ for
moderate synchronization ($C=0.2$, $D_1=0.6$) in dependence on the time delay $\tau$. Panels (a), (b), and (c)
correspond to a memory parameter of $R=0$, $R=0.5$, and $R=0.9$, respectively. The feedback strength is fixed at 
$K=1.5$. Other parameters as in Fig.~\ref{fig:x_t_mod_weak_strong}.}
\end{figure*}

\begin{figure*}[th]
  \begin{center}
	\includegraphics[width=\linewidth]{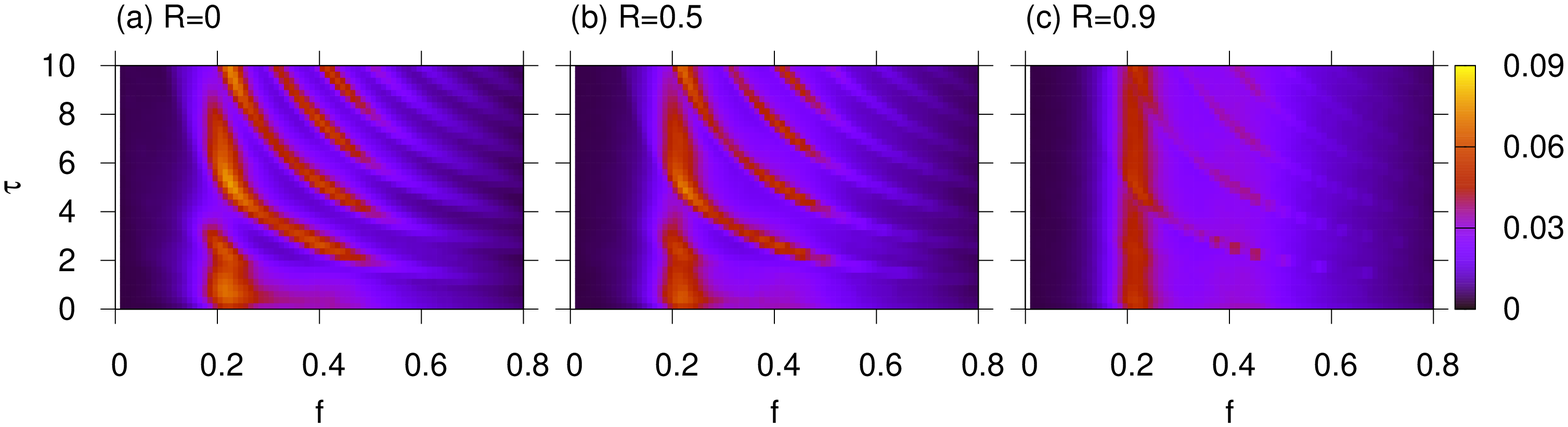}
  \end{center}
  \caption{\label{fig:spectrum_weak}(Color online) Power spectrum of the summarized signal $x_{\Sigma}=x_1+x_2$ for weak
synchronization ($C=0.1$, $D_1=0.6$) in dependence on the time delay $\tau$. Panels (a), (b), and (c)
correspond to a memory parameter of $R=0$, $R=0.5$, and $R=0.9$, respectively. Other parameters as in
Fig.~\ref{fig:spectrum_mod}.}
\end{figure*}

\begin{figure*}[th]
  \begin{center} 
	\includegraphics[width=\linewidth]{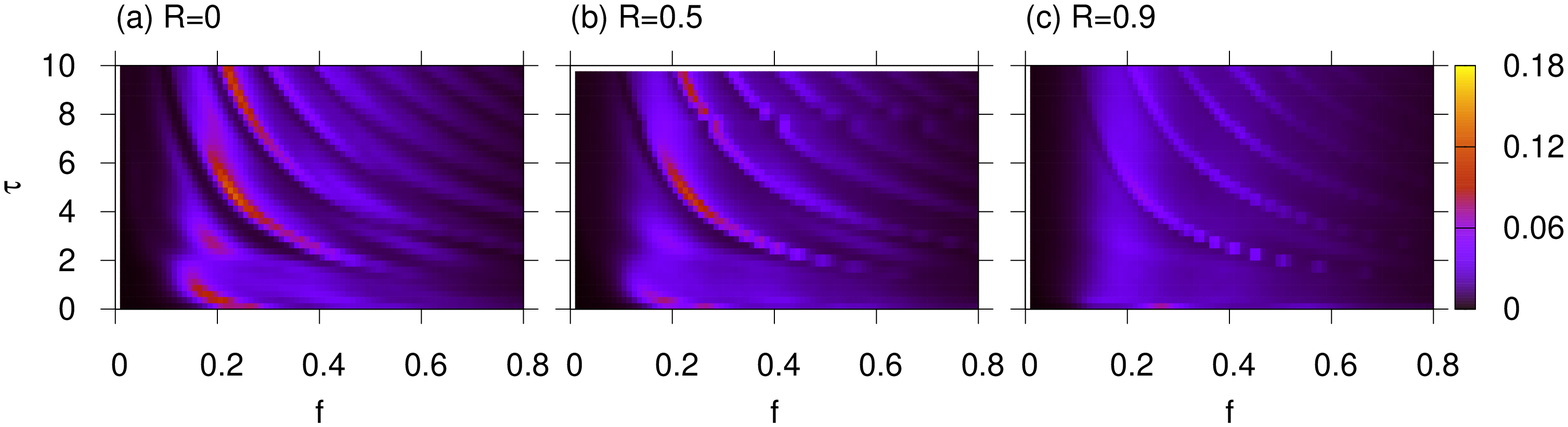}
  \end{center}
  \caption{\label{fig:spectrum_strong}(Color online) Power spectrum of the summarized signal $x_{\Sigma}=x_1+x_2$ for
strong synchronization ($C=0.2$, $D_1=0.15$) in dependence on the time delay $\tau$. Panels (a), (b), and (c)
correspond to a memory parameter of $R=0$, $R=0.5$, and $R=0.9$, respectively. Other parameters as in
Fig.~\ref{fig:spectrum_mod}.}
\end{figure*}

Next, we apply extended time-delayed feedback to the system. Figures~\ref{fig:spectrum_mod}, \ref{fig:spectrum_weak},
and \ref{fig:spectrum_strong} show the power spectrum of the summarized signal $x_\Sigma=x_1+x_2$ for the three above
mentioned cases of synchronization in dependence on the time delay $\tau$. In all figures, panels (a), (b), and (c)
correspond to a memory parameter of $R=0$, $R=0.5$, and $R=0.9$, respectively. The feedback gain is fixed
at $K=1.5$.

It can be seen that, depending on the choice of $\tau$, the main frequency component is shifted. Thus, the control
scheme is able to support different timescales. Note that, for instance in the case of moderate synchronization
(Fig.~\ref{fig:spectrum_mod}), the control force enhances the frequency corresponding to the dynamics of the $x_1$- or
$x_2$-variable. Compare the bright (yellow) areas in Fig.~\ref{fig:spectrum_mod} to the middle panel of the
uncontrolled case in Fig.~\ref{fig:spectrum_nocontrol}. A time delay of $\tau \approx 3$ favors a frequency of $f\approx
0.4$, i.e., the main frequency of $x_1$, and $\tau \approx 5$ enhances components of $f\approx 0.2$, which corresponds
to the dynamics of $x_2$.

For larger memory parameters $R$, this effect is less pronounced. The power spectra of the controlled system display
reduced sensitivity on the specific choice of the time delay. The main peak of the spectrum is stronger localized at
the frequency of the second subsystem, but the value of the power spectrum at this main frequency is much lower. 

\begin{figure*}[th]
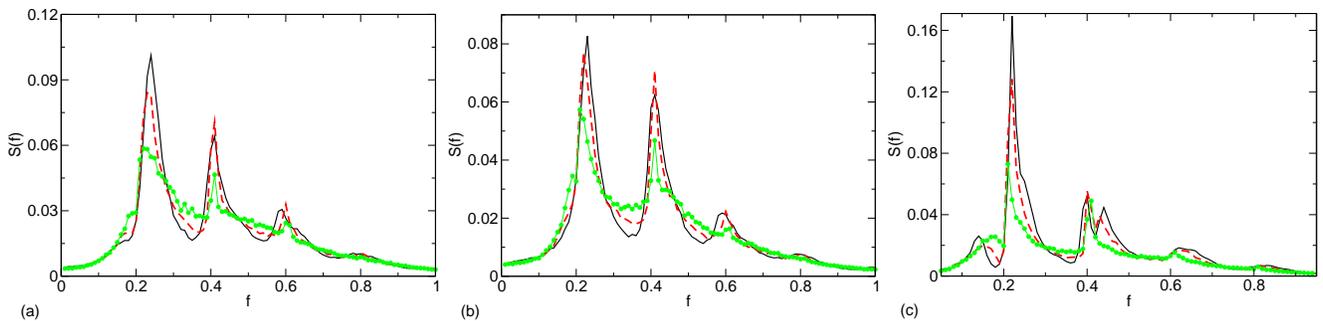

  \begin{center} 
	\includegraphics[width=0.32\linewidth]{figure10a.eps}
	\includegraphics[width=0.32\linewidth]{figure10b.eps}
	\includegraphics[width=0.32\linewidth]{figure10c.eps}
  \end{center}
  \caption{\label{fig:spectrum_mix_tau5}(Color online) Power spectrum of the summarized signal $x_{\Sigma}=x_1+x_2$ for
a fixed time delay of $\tau=5$. Panels (a), (b), and (c) correspond the case of moderate ($C=0.2$, $D_1=0.6$), weak
($C=0.1$, $D_1=0.6$), and strong ($C=0.2$, $D_1=0.15$) synchronization. The solid (black), dashed (red), and dotted
(green) curves refer to a memory parameter of $R=0$, $0.5$, and $0.9$, respectively. Other parameters as in
Fig.~\ref{fig:spectrum_mod}.}
\end{figure*}

Figure~\ref{fig:spectrum_mix_tau5} shows the power spectrum of the global summarized signal $x_\Sigma$ for the three
different case of synchronization and fixed time delay $\tau=5$. In all panels, the solid (black), dashed (red), and
and dotted (green) curves correspond to a memory parameter of $R=0$, $0.5$, and $0.9$, respectively. The feedback gain
is fixed at $K=1.5$. Panels (a), (b), and (c) of Fig.~\ref{fig:spectrum_mix_tau5} can be seen as horizontal cuts (at
$\tau=5$) of Figs.~\ref{fig:spectrum_mod}, \ref{fig:spectrum_weak}, and \ref{fig:spectrum_strong}, respectively.

Opposed to the uncontrolled case (see Fig.~\ref{fig:spectrum_nocontrol}), the power spectra exhibit a series of
distinct peaks that are located at the harmonics of the main frequency $f\approx 0.2$. They are due to the control
force which enhances not only the main frequency, but also frequency components of the higher harmonics. For increasing
memory parameter $R$, the background becomes broader. We stress that a similar effect was found in the context of
extended time-delayed feedback applied to the noise-induced oscillations in the Van-der-Pol system \citep{POM07}
and in a reaction-diffusion system \citep{MAJ09}, where,
in addition, the peaks become sharper.

\section{Interspike interval distribution}
\label{sec:distr} 
In the previous Section, we have discussed the modulation of timescales due to time-delayed feedback.
To obtain further information about the timescales present in the coupled system, we 
investigate the probability distribution of
the interspike intervals in the following, where we restrict our investigation to the analysis of the summarized signal
$x_{\Sigma}=x_1+x_2$.

\begin{figure*}[th]
  \begin{center} 
	\includegraphics[width=\linewidth]{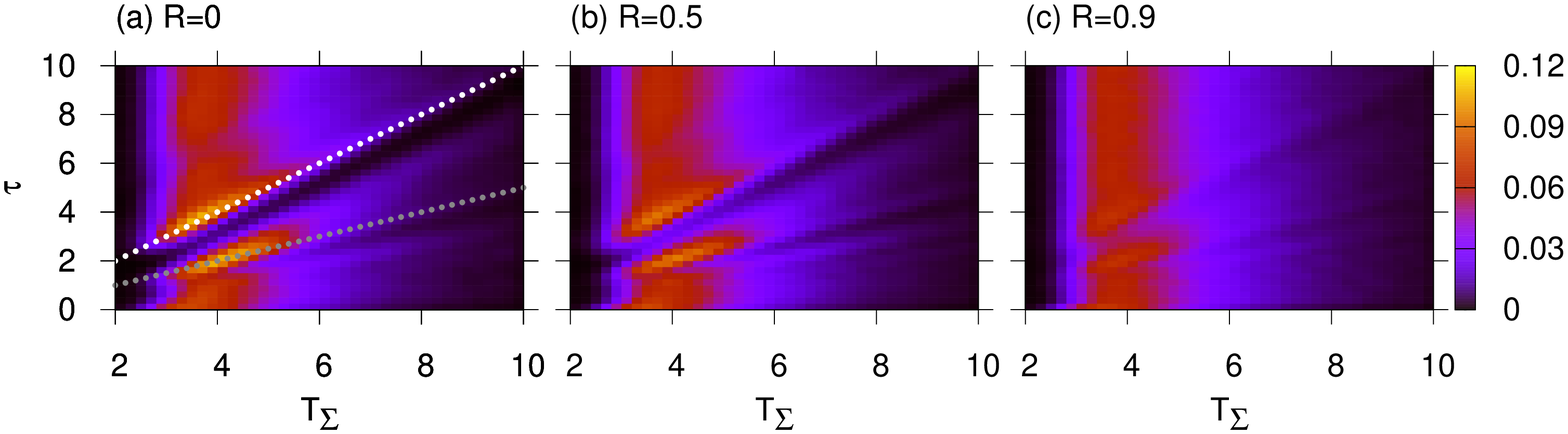}
  \end{center}
  \caption{\label{fig:distr_mod}(Color online) Interspike interval ($T_{\Sigma}$) distribution of the
summarized signal $x_{\Sigma}=x_1+x_2$ for moderate synchronization ($C=0.2$, $D_1=0.6$) in dependence on the time delay
$\tau$. The greyscale (color coding) denotes the probability of finding a certain interspike interval $T_{\Sigma}$.
Panels (a), (b), and (c) correspond to a memory parameter of $R=0$, $R=0.5$, and $R=0.9$, respectively. The white and
grey dotted lines in panel (a) at $\tau=T_{\Sigma}$ and $\tau=T_{\Sigma}/2$, respectively, are guides to the eye. Other
parameters as in Fig.~\ref{fig:spectrum_mod}.}
\end{figure*}

\begin{figure*}[th]
  \begin{center}
	\includegraphics[width=\linewidth]{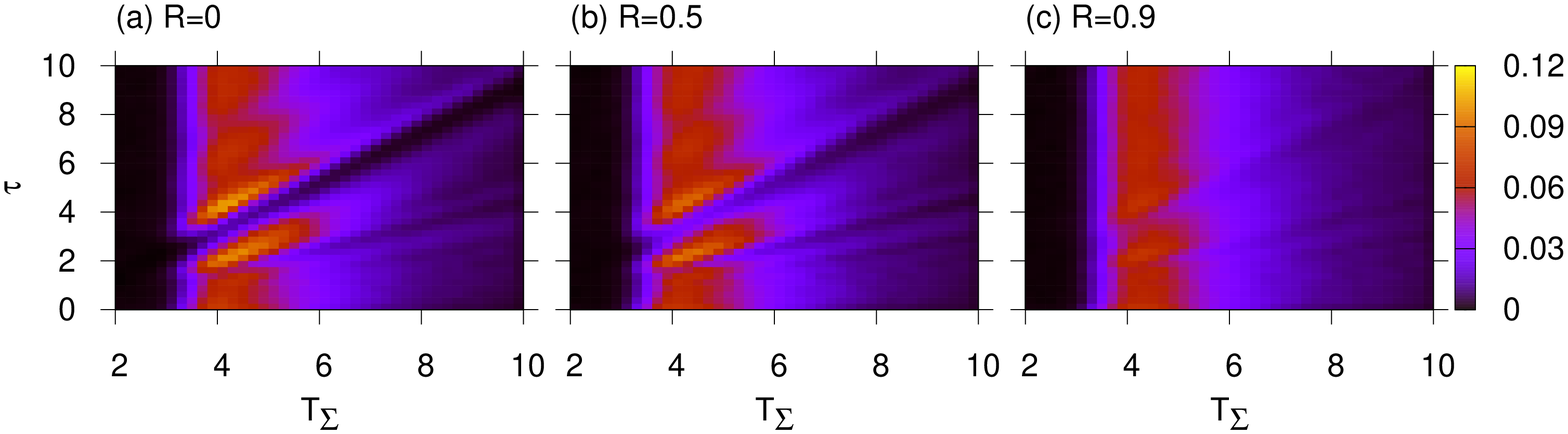}
  \end{center}
  \caption{\label{fig:distr_weak}(Color online) Interspike interval ($T_{\Sigma}$) distribution of the
summarized signal $x_{\Sigma}=x_1+x_2$ for weak synchronization ($C=0.1$, $D_1=0.6$) in dependence on the time delay
$\tau$. The greyscale (color coding) denotes the probability of finding a certain interspike interval $T_{\Sigma}$.
Panels (a), (b), and (c) correspond to a memory parameter of $R=0$, $R=0.5$, and $R=0.9$, respectively. Other parameters
as in Fig.~\ref{fig:spectrum_mod}.}
\end{figure*}

\begin{figure*}[th]
  \begin{center} 
	\includegraphics[width=\linewidth]{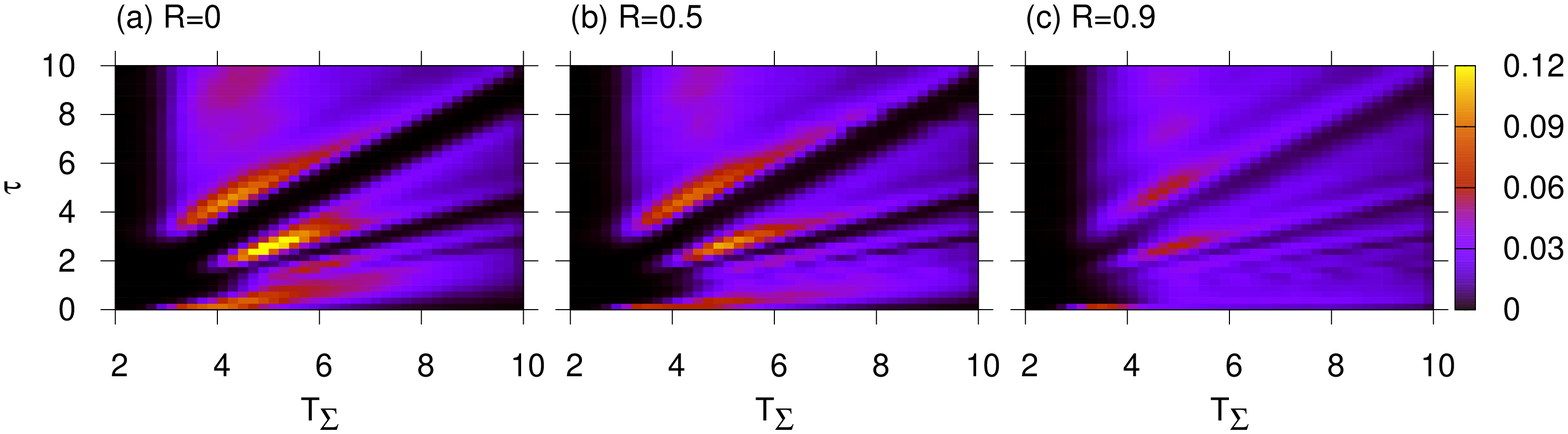}
  \end{center}
  \caption{\label{fig:distr_strong}(Color online) Interspike interval ($T_{\Sigma}$) distribution of the
summarized signal $x_{\Sigma}=x_1+x_2$ for strong synchronization ($C=0.2$, $D_1=0.15$) in dependence on the time delay
$\tau$. The greyscale (color coding) denotes the probability of finding a certain interspike interval $T_{\Sigma}$.
Panels (a), (b), and (c) correspond to a memory parameter of $R=0$, $R=0.5$, and $R=0.9$, respectively. Other parameters
as in Fig.~\ref{fig:spectrum_mod}.}
\end{figure*}

\begin{figure*}[th]
  \begin{center} 
	\includegraphics[width=\linewidth]{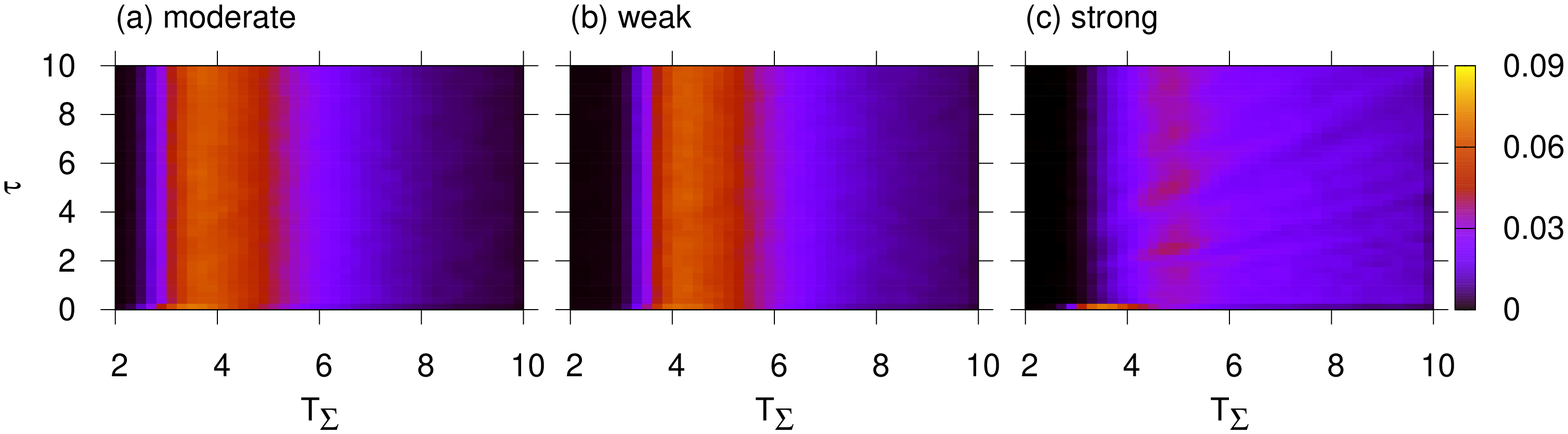}
  \end{center}
  \caption{\label{fig:distr_R0.99}(Color online) Interspike interval ($T_{\Sigma}$) distribution of the summarized
signal $x_{\Sigma}=x_1+x_2$ in dependence on the time delay $\tau$ for a memory parameter $R=0.99$ for moderate
($C=0.2$, $D_1=0.6$), weak ($C=0.1$, $D_1=0.6$), and strong ($C=0.2$, $D_1=0.15$) synchronization in panels (a), (b),
and (c), respectively. The greyscale (color coding) denotes the probability of finding a certain interspike interval
$T_{\Sigma}$. Other parameters as in Fig.~\ref{fig:spectrum_mod}.}
\end{figure*}

Figures~\ref{fig:distr_mod}, \ref{fig:distr_weak}, and \ref{fig:distr_strong} depict the dependence of
the distributions of the interspike interval $T_{\Sigma}$ for moderately ($C=0.2$, $D_1=0.6$), weakly ($C=0.1$,
$D_1=0.6$), and strongly synchronized ($C=0.2$, $D_1=0.15$) subsystems on the time delay, respectively. These are the
three cases marked in Fig.~\ref{fig:isi_D1_C}. The corresponding power spectra are shown in
Figs.~\ref{fig:spectrum_mod}, \ref{fig:spectrum_weak}, and \ref{fig:spectrum_strong}. Panels(a), (b), and (c) in each
figure refer to a memory parameter of $R=0$, $0.5$, and $0.9$, respectively. Note that all three
figures \ref{fig:distr_mod}, \ref{fig:distr_weak}, and \ref{fig:distr_strong} display the same greyscale (color code).

For increasing memory parameter $R$, the distribution exhibits a weaker dependence on the time delay
$\tau$. This can be seen, for instance, in panels (c) of Figs.~\ref{fig:distr_mod} and \ref{fig:distr_weak}. For smaller
$R$ values, another effect becomes apparent. There are a minima in the distribution for interspike intervals slightly
larger than $\tau$, indicated by black stripes in the $(T_{\Sigma},\tau)$ plane. These minima are located below the line $T_{\Sigma}= \tau$ (white line in
Fig.~\ref{fig:distr_mod}(a)). A similar, but less pronounced
effect can be observed if the interspike intervals match integer multiples of the time delay as shown by the grey line
for $T_{\Sigma}=2\tau$ in the same panel. We stress that for strongly synchronized subsystems this structuring of
the interspike interval distribution becomes more visible as displayed in Fig.~\ref{fig:distr_strong}. The probability
distribution becomes multimodal with peaks centered near $T_{\Sigma}=n\tau, n=1,2,\dots$.

To summarize this Section, the introduction of a large memory parameter renders the control method more
robust against the specific choice of the time delay as is depicted in Fig.~\ref{fig:distr_R0.99} for $R=0.99$. However, for small memory parameters, there is a competing
 effect which structures the distribution in the sense that interspike intervals slightly larger than the time delay of
the feedback are suppressed.

\section{Phase synchronization}
\label{sec:APS}

\begin{figure}[th]
  \begin{center}
	\includegraphics[width=\linewidth]{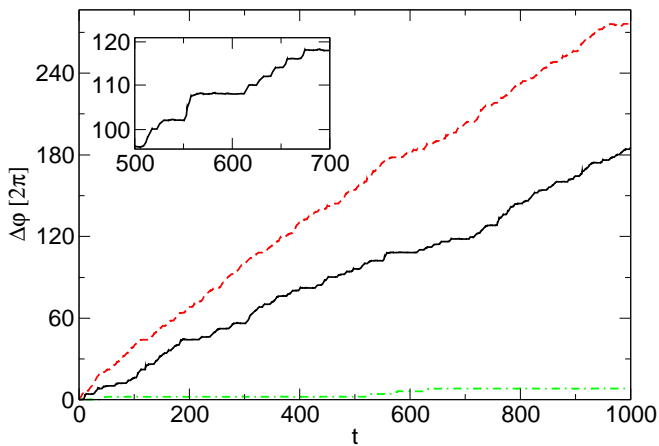}
  \end{center}
  \caption{\label{fig:phasendiff}(Color online) Phase difference in units of $2\pi$. The solid (black), dashed (red),
and dash-dotted (green) curves correspond to the case of moderate ($C=0.2$, $D_1=0.6$), weak ($C=0.1$, $D_1=0.6$), and
strong ($C=0.2$, $D_1=0.15$) synchronization, respectively. No control is applied to the system. The inset shows an
enlargement for moderate synchronization. Other parameters as in Fig.~\ref{fig:x_t_mod_weak_strong}.}
\end{figure}

The measures for cooperative dynamics considered so far are insensitive to phase relations. A measure of phase
synchronization can be obtained by introducing phase variables for each subsystems, and monitoring their difference. 

For definition of such a phase variable, one can generate a phase from the time series of spikes:
\begin{eqnarray}
 	\varphi(t) &=& 2\pi\frac{t-t_{i-1}}{t_{i}-t_{i-1}}+2\pi (i-1),
\end{eqnarray}
where $t_{i}$ denotes the time of the $i$-th spike. With this definition the phase increases by a value of
$2\pi$ for each spike \citep{PIK96,PIK01,HAU05,HAU06}. The phase difference $\Delta \varphi$ between two subsystems
can be defined for general $n:m$ synchronization as follows
\begin{eqnarray}
 \Delta\varphi_{n,m}(t) &=& |\varphi_{1}(t) - \frac{m}{n}\varphi_{2}(t)|,
\end{eqnarray} 
where $\varphi_{1}(t)$ and $\varphi_{2}(t)$ denote the phases of the respective subunits. In this work, we consider
only $1:1$-synchronization. 

In Refs.~\citep{ROS01,HAU06}, a measure for the phase synchronization is considered: the so-called
synchronization index $\gamma$. This quantity is also defined using the phase difference $\Delta \varphi$:
\begin{eqnarray} 
 \label{eq:gamma} 
 \gamma = \sqrt{\langle\cos \Delta \varphi(t)\rangle^2+\langle\sin \Delta \varphi(t)\rangle^2}.
\end{eqnarray}
It varies between $0$ (no synchronization) and $1$ (perfect synchronization). 

If the two subsystems are already strongly synchronized, it is helpful to consider the time intervals during which the
phase difference stays in a $2\pi$-phase range \citep{PAR05a,LAI06,PAR07}. In these synchronization intervals, the
subsystems exhibit the same number of spikes. In the case of strong synchronization, the synchronization index $\gamma$
shows only small changes near its maximum value, whereas the time interval of constant phase difference can vary
significantly. 

\begin{figure}[th]
  \begin{center}
	\includegraphics[width=\linewidth]{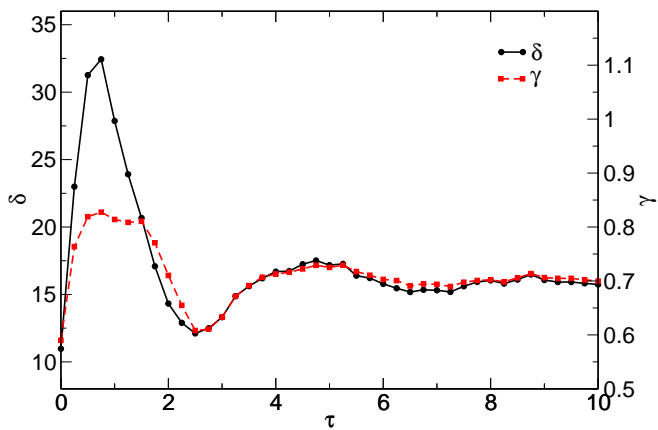}
  \end{center}
  \caption{\label{fig:delta_gamma}(Color online) Average phase synchronization interval $\delta$ (black dots, solid
curve) and synchronization index $\gamma$ (red squares, dashed line) in dependence on the time delay $\tau$ for the case
of moderately synchronized subsystems and vanishing memory parameter $R$. Other parameters as in
Fig.~\ref{fig:phasendiff}.}
\end{figure}

A measure of the amount of synchronization is given by the average length $\delta$ of the synchronization intervals. The
stronger sensitivity, as compared to $\gamma$, can be seen in Fig.~\ref{fig:delta_gamma}, which depicts the average
phase synchronization interval $\delta$ as black dots (solid curve) and the synchronization index $\gamma$ as red
squares (dashed line) for varying time delay $\tau$. The other control parameters are chosen as $R=0$ and $K=1.5$. The
scale is chosen such that the points for $\tau=0$ and $\tau=10$ coincide. Note that the average phase synchronization
interval $\delta$ shows a stronger increase than the synchronization index $\gamma$. Since the average phase
synchronization interval $\delta$ is more sensitive to the control parameters $K$, $\tau$, and $R$ than the
synchronization index $\gamma$, we restrict our investigations to the discussion of $\delta$.
 
\begin{figure*}[th]
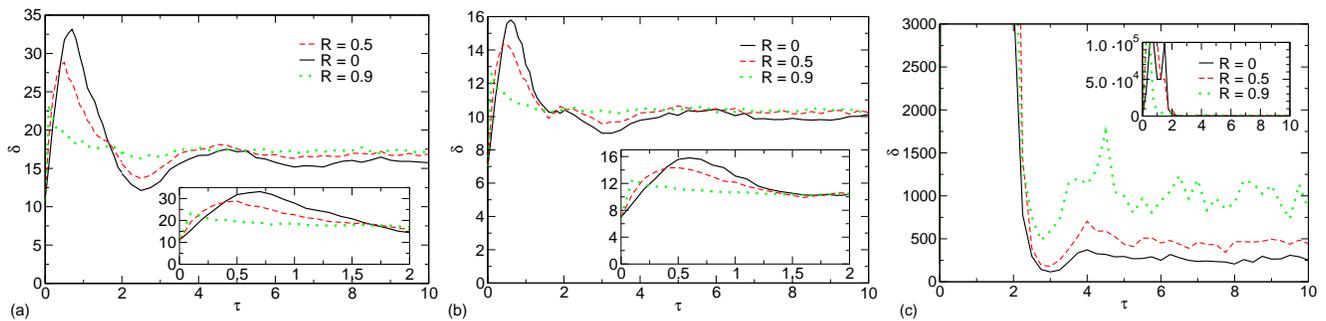

  \begin{center}
	\includegraphics[width=0.32\linewidth]{figure17a.eps}
	\includegraphics[width=0.32\linewidth]{figure17b.eps}
	\includegraphics[width=0.32\linewidth]{figure17c.eps}
  \end{center}
  \caption{\label{fig:aps}(Color online) Average phase synchronization interval $\delta$ in dependence on the time delay
$\tau$. The solid (black), dashed (red), and dotted (green) curves correspond to a memory parameter of $R=0$, $R=0.5$,
and $R=0.9$, respectively. The panels (a), (b), and (c) refer to the case of moderate, weak, and strong synchronization,
respectively. The insets in panels (a) and (b) show an enlargement for small $\tau$. The inset in panel (c) displays
 large $\delta$. Other parameters as in Fig.~\ref{fig:phasendiff}.}
\end{figure*}

Figure~\ref{fig:phasendiff} displays the time evolution of the phase difference $\Delta\varphi$, when no control is
applied to the system. The solid (black), dashed (red), and dash-dotted (green) curves correspond to the case of
moderate ($C=0.2$, $D_1=0.6$), weak ($C=0.1$, $D_1=0.6$), and strong ($C=0.2$, $D_1=0.15$) synchronization,
respectively. The inset depicts an enlargement for moderate synchronization. From this inset, one can clearly see the
plateaus in between phase jumps of $2\pi$. At these jumps, only one subsystem shows a spike whereas the other one
remains subthreshold. For better synchronization, the slope of $\Delta \varphi$ becomes flatter. In the case of strong
synchronization, for instance, $\Delta \varphi$ shows only a few phase jumps and remains in a $2\pi$-range for large
time intervals. The quantity $\delta$ measures the average lengths of these intervals.

Figure~\ref{fig:aps} shows effects of extended time-delayed feedback in the average phase synchronization interval
$\delta$ for varying time delay $\tau$. The feedback gain $K$ is fixed at $K=1.5$. Panels (a), (b), and (c) display the
case of moderately, weakly, and strongly synchronized subsystems, respectively. In all panels, the solid (black), dashed
(red), and dotted (green) curves refer to a memory parameter of $R=0$, $R=0.5$, and $R=0.9$, respectively. The insets in
panels (a) and (b) are enlargements for small time delays and the inset in panel (c) display large values of $\delta$.

In general, time-delayed feedback enlarges the average phase synchronization interval $\delta$. Especially for small
time delays, e.g., $\tau=0.7$ for $R=0$, $\delta$ becomes substantially larger. For $R=0$, a modulation of $\delta$ can
be seen for small delays, see insets in all panels of Fig.~\ref{fig:aps}. These deviations are less pronounced for
increasing $R$. Only panel (c), which refers to strong synchronization, shows larger values of $\delta$ with increasing
memory parameter. 

The sensitivity of $\delta$, as discussed in Fig.~\ref{fig:delta_gamma}, can also be seen in the case of strong
synchronization. See panel (c) in Fig.~\ref{fig:aps}. Since the already strong synchronization is further enhanced by
the control force, the average phase synchronization interval rises by several orders of magnitude as shown in the
inset. For perfect synchronization and simultaneous spiking, $\delta$ would be arbitrarily large and merely reflect the
integration time.

\section{Conclusion}
\label{sec:conclusion}
We have investigated the cooperative dynamics of two symmetrically coupled neurons under the effects of local extended
time-delayed feedback. We have found that the specific choice of the control parameters, i.e., feedback gain $K$, time
delay $\tau$, and memory parameter $R$, alters the cooperativity, which we have discussed in the context of different
measures of synchronization: ratio of average interspike intervals and average phase synchronization intervals. If the
control force is generated including states further in the past, i.e., for larger memory parameter, the two subsystems
exhibit enhanced phase synchronization, if the uncontrolled system is already in the strong synchronization regime. In
general, both the frequency and phase synchronization as well as the interspike interval distribution
become less sensitive to variations in $\tau$ with increasing
$R$. A small memory parameter $R$ leads to a suppression of interspike intervals slightly larger than
the time delay.

The stability of a synchronous manifold for coupled ordinary differential equations can be
investigated by means of a master stability function \citep{PEC98,CHA05}. For delay differential equations
involving stochastic input, this formalism is not yet developed. Future investigations should address this to obtain
an analytical understanding of the control of synchronization in neural networks by time-delayed feedback. 
\begin{acknowledgments}
This work was supported by DFG in the framework of Sfb 555 (Complex Nonlinear Processes). P.~H. acknowledges
support of the Deutsche Akademische Austausch\-dienst (DAAD) and thanks Kazuyuki Aihara and his group for stimulating
discussions.
\end{acknowledgments}


\begin{thebibliography}{39}
\expandafter\ifx\csname natexlab\endcsname\relax\def\natexlab#1{#1}\fi
\expandafter\ifx\csname bibnamefont\endcsname\relax
  \def\bibnamefont#1{#1}\fi
\expandafter\ifx\csname bibfnamefont\endcsname\relax
  \def\bibfnamefont#1{#1}\fi
\expandafter\ifx\csname citenamefont\endcsname\relax
  \def\citenamefont#1{#1}\fi
\expandafter\ifx\csname url\endcsname\relax
  \def\url#1{\texttt{#1}}\fi
\expandafter\ifx\csname urlprefix\endcsname\relax\def\urlprefix{URL }\fi
\providecommand{\bibinfo}[2]{#2}
\providecommand{\eprint}[2][]{\url{#2}}
\hangafter=1
\hangindent=1.25cm

\bibitem[{\citenamefont{Balanov} \emph{et~al.}(2006)\citenamefont{Balanov,
  Beato, Janson, Engel, \& Sch{\"o}ll}}]{BAL06}
\bibinfo{author}{\bibnamefont{Balanov}, \bibfnamefont{A.~G.}},
  \bibinfo{author}{\bibnamefont{Beato}, \bibfnamefont{V.}},
  \bibinfo{author}{\bibnamefont{Janson}, \bibfnamefont{N.~B.}},
  \bibinfo{author}{\bibnamefont{Engel}, \bibfnamefont{H.}} \&
  \bibinfo{author}{\bibnamefont{Sch{\"o}ll}, \bibfnamefont{E.}}
  [\bibinfo{year}{2006}] ``\bibinfo{title}{Delayed feedback control of
  noise-induced patterns in excitable media},''
  \emph{\bibinfo{journal}{Phys.~Rev.~E}} \textbf{\bibinfo{volume}{74}},
  \bibinfo{pages}{016214}.

\bibitem[{\citenamefont{Balanov} \emph{et~al.}(2004)\citenamefont{Balanov,
  Janson, \& Sch{\"o}ll}}]{BAL04}
\bibinfo{author}{\bibnamefont{Balanov}, \bibfnamefont{A.~G.}},
  \bibinfo{author}{\bibnamefont{Janson}, \bibfnamefont{N.~B.}} \&
  \bibinfo{author}{\bibnamefont{Sch{\"o}ll}, \bibfnamefont{E.}}
  [\bibinfo{year}{2004}] ``\bibinfo{title}{Control of noise-induced
  oscillations by delayed feedback},'' \emph{\bibinfo{journal}{Physica~D}}
  \textbf{\bibinfo{volume}{199}}, \bibinfo{pages}{1}.

\bibitem[{\citenamefont{Chavez} \emph{et~al.}(2005)\citenamefont{Chavez, Hwang,
  Amann, Hentschel, \& Boccaletti}}]{CHA05}
\bibinfo{author}{\bibnamefont{Chavez}, \bibfnamefont{M.}},
  \bibinfo{author}{\bibnamefont{Hwang}, \bibfnamefont{D.~U.}},
  \bibinfo{author}{\bibnamefont{Amann}, \bibfnamefont{A.}},
  \bibinfo{author}{\bibnamefont{Hentschel}, \bibfnamefont{H.~G.~E.}} \&
  \bibinfo{author}{\bibnamefont{Boccaletti}, \bibfnamefont{S.}}
  [\bibinfo{year}{2005}] ``\bibinfo{title}{Synchronization is Enhanced in
  Weighted Complex Networks},'' \emph{\bibinfo{journal}{Phys.~Rev.~Lett.}}
  \textbf{\bibinfo{volume}{94}}, \bibinfo{pages}{218701}.

\bibitem[{\citenamefont{Dahms} \emph{et~al.}(2007)\citenamefont{Dahms,
  H{\"o}vel, \& Sch{\"o}ll}}]{DAH07}
\bibinfo{author}{\bibnamefont{Dahms}, \bibfnamefont{T.}},
  \bibinfo{author}{\bibnamefont{H{\"o}vel}, \bibfnamefont{P.}} \&
  \bibinfo{author}{\bibnamefont{Sch{\"o}ll}, \bibfnamefont{E.}}
  [\bibinfo{year}{2007}] ``\bibinfo{title}{Control of unstable steady states by
  extended time-delayed feedback},'' \emph{\bibinfo{journal}{Phys.~Rev.~E}}
  \textbf{\bibinfo{volume}{76}}(\bibinfo{number}{5}), \bibinfo{pages}{056201}.

\bibitem[{\citenamefont{Dahms} \emph{et~al.}(2008)\citenamefont{Dahms,
  H{\"o}vel, \& Sch{\"o}ll}}]{DAH08b}
\bibinfo{author}{\bibnamefont{Dahms}, \bibfnamefont{T.}},
  \bibinfo{author}{\bibnamefont{H{\"o}vel}, \bibfnamefont{P.}} \&
  \bibinfo{author}{\bibnamefont{Sch{\"o}ll}, \bibfnamefont{E.}}
  [\bibinfo{year}{2008}] ``\bibinfo{title}{Stabilizing continuous-wave output
  in semiconductor lasers by time-delayed feedback},''
  \emph{\bibinfo{journal}{Phys. Rev. E}} \bibinfo{note}{Submitted}.

\bibitem[{\citenamefont{Flunkert \& Sch{\"o}ll}(2007)}]{FLU07}
\bibinfo{author}{\bibnamefont{Flunkert}, \bibfnamefont{V.}} \&
  \bibinfo{author}{\bibnamefont{Sch{\"o}ll}, \bibfnamefont{E.}}
  [\bibinfo{year}{2007}] ``\bibinfo{title}{Suppressing noise-induced intensity
  pulsations in semiconductor lasers by means of time-delayed feedback},''
  \emph{\bibinfo{journal}{Phys. Rev. E}} \textbf{\bibinfo{volume}{76}},
  \bibinfo{pages}{066202}.

\bibitem[{\citenamefont{Gassel} \emph{et~al.}(2007)\citenamefont{Gassel, Glatt,
  \& Kaiser}}]{GAS07b}
\bibinfo{author}{\bibnamefont{Gassel}, \bibfnamefont{M.}},
  \bibinfo{author}{\bibnamefont{Glatt}, \bibfnamefont{E.}} \&
  \bibinfo{author}{\bibnamefont{Kaiser}, \bibfnamefont{F.}}
  [\bibinfo{year}{2007}] ``\bibinfo{title}{Time-delayed feedback in a net of
  neural elements: Transitions from oscillatory to excitable dynamics},''
  \emph{\bibinfo{journal}{Fluct. Noise Lett.}}
  \textbf{\bibinfo{volume}{7}}(\bibinfo{number}{3}), \bibinfo{pages}{L225}.

\bibitem[{\citenamefont{Gassel} \emph{et~al.}(2008)\citenamefont{Gassel, Glatt,
  \& Kaiser}}]{GAS08}
\bibinfo{author}{\bibnamefont{Gassel}, \bibfnamefont{M.}},
  \bibinfo{author}{\bibnamefont{Glatt}, \bibfnamefont{E.}} \&
  \bibinfo{author}{\bibnamefont{Kaiser}, \bibfnamefont{F.}}
  [\bibinfo{year}{2008}] ``\bibinfo{title}{Delay-sustained pattern formation in
  subexcitable media},'' \emph{\bibinfo{journal}{Phys.~Rev.~E}}
  \textbf{\bibinfo{volume}{77}}(\bibinfo{number}{6}), \bibinfo{eid}{066220}
  (pages~\bibinfo{numpages}{7}).

\bibitem[{\citenamefont{Hauschildt}(2005)}]{HAU05}
\bibinfo{author}{\bibnamefont{Hauschildt}, \bibfnamefont{B.}}
  [\bibinfo{year}{2005}] \emph{\bibinfo{title}{Control of noise-induced
  multimode oscillations in coupled neural systems}} Master's thesis
  \bibinfo{school}{TU Berlin}.

\bibitem[{\citenamefont{Hauschildt}
  \emph{et~al.}(2006)\citenamefont{Hauschildt, Janson, Balanov, \&
  Sch{\"o}ll}}]{HAU06}
\bibinfo{author}{\bibnamefont{Hauschildt}, \bibfnamefont{B.}},
  \bibinfo{author}{\bibnamefont{Janson}, \bibfnamefont{N.~B.}},
  \bibinfo{author}{\bibnamefont{Balanov}, \bibfnamefont{A.~G.}} \&
  \bibinfo{author}{\bibnamefont{Sch{\"o}ll}, \bibfnamefont{E.}}
  [\bibinfo{year}{2006}] ``\bibinfo{title}{Noise-induced cooperative dynamics
  and its control in coupled neuron models},''
  \emph{\bibinfo{journal}{Phys.~Rev.~E}} \textbf{\bibinfo{volume}{74}},
  \bibinfo{pages}{051906}.

\bibitem[{\citenamefont{Hizanidis} \emph{et~al.}(2006)\citenamefont{Hizanidis,
  Balanov, Amann, \& Sch{\"o}ll}}]{HIZ06}
\bibinfo{author}{\bibnamefont{Hizanidis}, \bibfnamefont{J.}},
  \bibinfo{author}{\bibnamefont{Balanov}, \bibfnamefont{A.~G.}},
  \bibinfo{author}{\bibnamefont{Amann}, \bibfnamefont{A.}} \&
  \bibinfo{author}{\bibnamefont{Sch{\"o}ll}, \bibfnamefont{E.}}
  [\bibinfo{year}{2006}] ``\bibinfo{title}{Noise-induced front motion:
  signature of a global bifurcation},''
  \emph{\bibinfo{journal}{Phys.~Rev.~Lett.}} \textbf{\bibinfo{volume}{96}},
  \bibinfo{pages}{244104}.

\bibitem[{\citenamefont{Hizanidis \& Sch{\"o}ll}(2008)}]{HIZ08}
\bibinfo{author}{\bibnamefont{Hizanidis}, \bibfnamefont{J.}} \&
  \bibinfo{author}{\bibnamefont{Sch{\"o}ll}, \bibfnamefont{E.}}
  [\bibinfo{year}{2008}] ``\bibinfo{title}{Control of noise-induced
  spatiotemporal patterns in superlattices},''
  \emph{\bibinfo{journal}{phys.~stat.~sol.~(c)}}
  \textbf{\bibinfo{volume}{5}}(\bibinfo{number}{1}), \bibinfo{pages}{207}.

\bibitem[{\citenamefont{Janson} \emph{et~al.}(2004)\citenamefont{Janson,
  Balanov, \& Sch{\"o}ll}}]{JAN03}
\bibinfo{author}{\bibnamefont{Janson}, \bibfnamefont{N.~B.}},
  \bibinfo{author}{\bibnamefont{Balanov}, \bibfnamefont{A.~G.}} \&
  \bibinfo{author}{\bibnamefont{Sch{\"o}ll}, \bibfnamefont{E.}}
  [\bibinfo{year}{2004}] ``\bibinfo{title}{Delayed Feedback as a Means of
  Control of Noise-Induced Motion},''
  \emph{\bibinfo{journal}{Phys.~Rev.~Lett.}} \textbf{\bibinfo{volume}{93}},
  \bibinfo{pages}{010601}.

\bibitem[{\citenamefont{Lai} \emph{et~al.}(2006)\citenamefont{Lai, Frei, \&
  Osorio}}]{LAI06}
\bibinfo{author}{\bibnamefont{Lai}, \bibfnamefont{Y.~C.}},
  \bibinfo{author}{\bibnamefont{Frei}, \bibfnamefont{M.~G.}} \&
  \bibinfo{author}{\bibnamefont{Osorio}, \bibfnamefont{I.}}
  [\bibinfo{year}{2006}] ``\bibinfo{title}{{Detecting and characterizing phase
  synchronization in nonstationary dynamical systems}},''
  \emph{\bibinfo{journal}{Phys.~Rev.~E}}
  \textbf{\bibinfo{volume}{73}}(\bibinfo{number}{2}), \bibinfo{pages}{26214}.

\bibitem[{\citenamefont{Majer \& Sch{\"o}ll}(2009)}]{MAJ09}
\bibinfo{author}{\bibnamefont{Majer}, \bibfnamefont{N.}} \&
  \bibinfo{author}{\bibnamefont{Sch{\"o}ll}, \bibfnamefont{E.}}
  [\bibinfo{year}{2009}] ``\bibinfo{title}{Resonant control of stochastic
  spatio-temporal dynamics in a tunnel diode by multiple time delayed
  feedback},'' \emph{\bibinfo{journal}{Phys.~Rev.~E}} .

\bibitem[{\citenamefont{Park \& Lai}(2005)}]{PAR05a}
\bibinfo{author}{\bibnamefont{Park}, \bibfnamefont{K.}} \&
  \bibinfo{author}{\bibnamefont{Lai}, \bibfnamefont{Y.~C.}}
  [\bibinfo{year}{2005}] ``\bibinfo{title}{Characterization of stochastic
  resonance},'' \emph{\bibinfo{journal}{Europhys. Lett}}
  \textbf{\bibinfo{volume}{70}}(\bibinfo{number}{4}), \bibinfo{pages}{432}.

\bibitem[{\citenamefont{Park} \emph{et~al.}(2007)\citenamefont{Park, Lai, \&
  Krishnamoorthy}}]{PAR07}
\bibinfo{author}{\bibnamefont{Park}, \bibfnamefont{K.}},
  \bibinfo{author}{\bibnamefont{Lai}, \bibfnamefont{Y.~C.}} \&
  \bibinfo{author}{\bibnamefont{Krishnamoorthy}, \bibfnamefont{S.}}
  [\bibinfo{year}{2007}] ``\bibinfo{title}{{Noise sensitivity of
  phase-synchronization time in stochastic resonance: Theory and
  experiment}},'' \emph{\bibinfo{journal}{Phys.~Rev.~E}}
  \textbf{\bibinfo{volume}{75}}(\bibinfo{number}{4}), \bibinfo{pages}{46205}.

\bibitem[{\citenamefont{Pecora \& Carroll}(1998)}]{PEC98}
\bibinfo{author}{\bibnamefont{Pecora}, \bibfnamefont{L.~M.}} \&
  \bibinfo{author}{\bibnamefont{Carroll}, \bibfnamefont{T.~L.}}
  [\bibinfo{year}{1998}] ``\bibinfo{title}{Master Stability Functions for
  Synchronized Coupled Systems},'' \emph{\bibinfo{journal}{Phys. Rev. Lett.}}
  \textbf{\bibinfo{volume}{80}}(\bibinfo{number}{10}), \bibinfo{pages}{2109}.

\bibitem[{\citenamefont{Pikovsky} \emph{et~al.}(1996)\citenamefont{Pikovsky,
  Rosenblum, \& Kurths}}]{PIK96}
\bibinfo{author}{\bibnamefont{Pikovsky}, \bibfnamefont{A.}},
  \bibinfo{author}{\bibnamefont{Rosenblum}, \bibfnamefont{M.~G.}} \&
  \bibinfo{author}{\bibnamefont{Kurths}, \bibfnamefont{J.}}
  [\bibinfo{year}{1996}] ``\bibinfo{title}{Synchronisation in a population of
  globally coupled chaotic oscillators},''
  \emph{\bibinfo{journal}{Europhys.~Lett.}} \textbf{\bibinfo{volume}{34}},
  \bibinfo{pages}{165}.

\bibitem[{\citenamefont{Pikovsky} \emph{et~al.}(2001)\citenamefont{Pikovsky,
  Rosenblum, \& Kurths}}]{PIK01}
\bibinfo{author}{\bibnamefont{Pikovsky}, \bibfnamefont{A.}},
  \bibinfo{author}{\bibnamefont{Rosenblum}, \bibfnamefont{M.~G.}} \&
  \bibinfo{author}{\bibnamefont{Kurths}, \bibfnamefont{J.}}
  [\bibinfo{year}{2001}] \emph{\bibinfo{title}{Synchronization, A Universal
  Concept in Nonlinear Sciences}} (\bibinfo{publisher}{Cambridge University
  Press}, \bibinfo{address}{Cambridge}).

\bibitem[{\citenamefont{Pomplun} \emph{et~al.}(2005)\citenamefont{Pomplun,
  Amann, \& Sch{\"o}ll}}]{POM05a}
\bibinfo{author}{\bibnamefont{Pomplun}, \bibfnamefont{J.}},
  \bibinfo{author}{\bibnamefont{Amann}, \bibfnamefont{A.}} \&
  \bibinfo{author}{\bibnamefont{Sch{\"o}ll}, \bibfnamefont{E.}}
  [\bibinfo{year}{2005}] ``\bibinfo{title}{Mean field approximation of
  time-delayed feedback control of noise-induced oscillations in the {V}an der
  {P}ol system},'' \emph{\bibinfo{journal}{Europhys.~Lett.}}
  \textbf{\bibinfo{volume}{71}}, \bibinfo{pages}{366}.

\bibitem[{\citenamefont{Pomplun} \emph{et~al.}(2007)\citenamefont{Pomplun,
  Balanov, \& Sch{\"o}ll}}]{POM07}
\bibinfo{author}{\bibnamefont{Pomplun}, \bibfnamefont{J.}},
  \bibinfo{author}{\bibnamefont{Balanov}, \bibfnamefont{A.~G.}} \&
  \bibinfo{author}{\bibnamefont{Sch{\"o}ll}, \bibfnamefont{E.}}
  [\bibinfo{year}{2007}] ``\bibinfo{title}{Long-term correlations in stochastic
  systems with extended time-delayed feedback},''
  \emph{\bibinfo{journal}{Phys.~Rev.~E}} \textbf{\bibinfo{volume}{75}},
  \bibinfo{pages}{040101(R)}.

\bibitem[{\citenamefont{Popovych} \emph{et~al.}(2005)\citenamefont{Popovych,
  Hauptmann, \& Tass}}]{POP05}
\bibinfo{author}{\bibnamefont{Popovych}, \bibfnamefont{O.~V.}},
  \bibinfo{author}{\bibnamefont{Hauptmann}, \bibfnamefont{C.}} \&
  \bibinfo{author}{\bibnamefont{Tass}, \bibfnamefont{P.~A.}}
  [\bibinfo{year}{2005}] ``\bibinfo{title}{Effective Desynchronization by
  Nonlinear Delayed Feedback},'' \emph{\bibinfo{journal}{Phys.~Rev.~Lett.}}
  \textbf{\bibinfo{volume}{94}}, \bibinfo{pages}{164102}.

\bibitem[{\citenamefont{Popovych} \emph{et~al.}(2006)\citenamefont{Popovych,
  Hauptmann, \& Tass}}]{POP06}
\bibinfo{author}{\bibnamefont{Popovych}, \bibfnamefont{O.~V.}},
  \bibinfo{author}{\bibnamefont{Hauptmann}, \bibfnamefont{C.}} \&
  \bibinfo{author}{\bibnamefont{Tass}, \bibfnamefont{P.~A.}}
  [\bibinfo{year}{2006}] ``\bibinfo{title}{Control of neuronal synchrony by
  nonlinear delayed feedback},'' \emph{\bibinfo{journal}{Biol. Cybern.}}
  \textbf{\bibinfo{volume}{95}}(\bibinfo{number}{1}), \bibinfo{pages}{69}.

\bibitem[{\citenamefont{Pototsky \& Janson}(2007)}]{POT07}
\bibinfo{author}{\bibnamefont{Pototsky}, \bibfnamefont{A.}} \&
  \bibinfo{author}{\bibnamefont{Janson}, \bibfnamefont{N.~B.}}
  [\bibinfo{year}{2007}] ``\bibinfo{title}{Correlation theory of delayed
  feedback in stochastic systems below Andronov-Hopf bifurcation},''
  \emph{\bibinfo{journal}{Phys.~Rev.~E}} \textbf{\bibinfo{volume}{76}},
  \bibinfo{pages}{056208}.

\bibitem[{\citenamefont{Pototsky \& Janson}(2008)}]{POT08}
\bibinfo{author}{\bibnamefont{Pototsky}, \bibfnamefont{A.}} \&
  \bibinfo{author}{\bibnamefont{Janson}, \bibfnamefont{N.~B.}}
  [\bibinfo{year}{2008}] ``\bibinfo{title}{Excitable systems with noise and
  delay, with applications to control: Renewal theory approach},''
  \emph{\bibinfo{journal}{Phys.~Rev.~E}}
  \textbf{\bibinfo{volume}{77}}(\bibinfo{number}{3}), \bibinfo{eid}{031113}
  (pages~\bibinfo{numpages}{11}).

\bibitem[{\citenamefont{Prager} \emph{et~al.}(2007)\citenamefont{Prager, Lerch,
  Schimansky-Geier, \& Sch{\"o}ll}}]{PRA07}
\bibinfo{author}{\bibnamefont{Prager}, \bibfnamefont{T.}},
  \bibinfo{author}{\bibnamefont{Lerch}, \bibfnamefont{H.~P.}},
  \bibinfo{author}{\bibnamefont{Schimansky-Geier}, \bibfnamefont{L.}} \&
  \bibinfo{author}{\bibnamefont{Sch{\"o}ll}, \bibfnamefont{E.}}
  [\bibinfo{year}{2007}] ``\bibinfo{title}{Increase of Coherence in Excitable
  Systems by Delayed Feedback},'' \emph{\bibinfo{journal}{J. Phys. A}}
  \textbf{\bibinfo{volume}{40}}, \bibinfo{pages}{11045}.

\bibitem[{\citenamefont{Pyragas}(1992)}]{PYR92}
\bibinfo{author}{\bibnamefont{Pyragas}, \bibfnamefont{K.}}
  [\bibinfo{year}{1992}] ``\bibinfo{title}{Continuous control of chaos by
  self-controlling feedback},'' \emph{\bibinfo{journal}{Phys.~Lett.~A}}
  \textbf{\bibinfo{volume}{170}}, \bibinfo{pages}{421}.

\bibitem[{\citenamefont{Rosenblum} \emph{et~al.}(2001)\citenamefont{Rosenblum,
  Pikovsky, Kurths, Sch{\"a}fer, \& Tass}}]{ROS01}
\bibinfo{author}{\bibnamefont{Rosenblum}, \bibfnamefont{M.}},
  \bibinfo{author}{\bibnamefont{Pikovsky}, \bibfnamefont{A.}},
  \bibinfo{author}{\bibnamefont{Kurths}, \bibfnamefont{J.}},
  \bibinfo{author}{\bibnamefont{Sch{\"a}fer}, \bibfnamefont{C.}} \&
  \bibinfo{author}{\bibnamefont{Tass}, \bibfnamefont{P.~A.}}
  [\bibinfo{year}{2001}] \emph{\bibinfo{title}{{Phase synchronization: from
  theory to data analysis}}} (\bibinfo{publisher}{Elsevier Science},
  \bibinfo{address}{Amsterdam}) volume~\bibinfo{volume}{4} of
  \emph{\bibinfo{series}{{Handbook of Biological Physics}}}
  chapter~\bibinfo{chapter}{9} pp. \bibinfo{pages}{279--321}.

\bibitem[{\citenamefont{Rosenblum \& Pikovsky}(2004{\natexlab{a}})}]{ROS04a}
\bibinfo{author}{\bibnamefont{Rosenblum}, \bibfnamefont{M.~G.}} \&
  \bibinfo{author}{\bibnamefont{Pikovsky}, \bibfnamefont{A.}}
  [\bibinfo{year}{2004}{\natexlab{a}}] ``\bibinfo{title}{Controlling
  Synchronization in an Ensemble of Globally Coupled Oscillators},''
  \emph{\bibinfo{journal}{Phys.~Rev.~Lett.}} \textbf{\bibinfo{volume}{92}},
  \bibinfo{pages}{114102}.

\bibitem[{\citenamefont{Rosenblum \& Pikovsky}(2004{\natexlab{b}})}]{ROS04}
\bibinfo{author}{\bibnamefont{Rosenblum}, \bibfnamefont{M.~G.}} \&
  \bibinfo{author}{\bibnamefont{Pikovsky}, \bibfnamefont{A.}}
  [\bibinfo{year}{2004}{\natexlab{b}}] ``\bibinfo{title}{Delayed feedback
  control of collective synchrony: An approach to suppression of pathological
  brain rhythms},'' \emph{\bibinfo{journal}{Phys.~Rev.~E}}
  \textbf{\bibinfo{volume}{70}}, \bibinfo{pages}{041904}.

\bibitem[{\citenamefont{Schiff} \emph{et~al.}(1994)\citenamefont{Schiff,
  Jerger, Duong, Chang, Spano, \& Ditto}}]{SCH94e}
\bibinfo{author}{\bibnamefont{Schiff}, \bibfnamefont{S.~J.}},
  \bibinfo{author}{\bibnamefont{Jerger}, \bibfnamefont{K.}},
  \bibinfo{author}{\bibnamefont{Duong}, \bibfnamefont{D.~H.}},
  \bibinfo{author}{\bibnamefont{Chang}, \bibfnamefont{T.}},
  \bibinfo{author}{\bibnamefont{Spano}, \bibfnamefont{M.~L.}} \&
  \bibinfo{author}{\bibnamefont{Ditto}, \bibfnamefont{W.~L.}}
  [\bibinfo{year}{1994}] ``\bibinfo{title}{Controlling Chaos in the brain},''
  \emph{\bibinfo{journal}{Nature (London)}} \textbf{\bibinfo{volume}{370}},
  \bibinfo{pages}{615}.

\bibitem[{\citenamefont{Schlesner} \emph{et~al.}(2003)\citenamefont{Schlesner,
  Amann, Janson, Just, \& Sch{\"o}ll}}]{SCH03a}
\bibinfo{author}{\bibnamefont{Schlesner}, \bibfnamefont{J.}},
  \bibinfo{author}{\bibnamefont{Amann}, \bibfnamefont{A.}},
  \bibinfo{author}{\bibnamefont{Janson}, \bibfnamefont{N.~B.}},
  \bibinfo{author}{\bibnamefont{Just}, \bibfnamefont{W.}} \&
  \bibinfo{author}{\bibnamefont{Sch{\"o}ll}, \bibfnamefont{E.}}
  [\bibinfo{year}{2003}] ``\bibinfo{title}{Self-stabilization of high frequency
  oscillations in semiconductor superlattices by time--delay
  autosynchronization},'' \emph{\bibinfo{journal}{Phys.~Rev.~E}}
  \textbf{\bibinfo{volume}{68}}, \bibinfo{pages}{066208}.

\bibitem[{\citenamefont{Sch{\"o}ll}
  \emph{et~al.}(2005)\citenamefont{Sch{\"o}ll, Balanov, Janson, \&
  Neiman}}]{SCH04b}
\bibinfo{author}{\bibnamefont{Sch{\"o}ll}, \bibfnamefont{E.}},
  \bibinfo{author}{\bibnamefont{Balanov}, \bibfnamefont{A.~G.}},
  \bibinfo{author}{\bibnamefont{Janson}, \bibfnamefont{N.~B.}} \&
  \bibinfo{author}{\bibnamefont{Neiman}, \bibfnamefont{A.}}
  [\bibinfo{year}{2005}] ``\bibinfo{title}{Controlling stochastic oscillations
  close to a {Hopf} bifurcation by time-delayed feedback},''
  \emph{\bibinfo{journal}{Stoch.~Dyn.}} \textbf{\bibinfo{volume}{5}},
  \bibinfo{pages}{281}.

\bibitem[{\citenamefont{Sch{\"o}ll}
  \emph{et~al.}(2008)\citenamefont{Sch{\"o}ll, Majer, \& Stegemann}}]{SCH08a}
\bibinfo{author}{\bibnamefont{Sch{\"o}ll}, \bibfnamefont{E.}},
  \bibinfo{author}{\bibnamefont{Majer}, \bibfnamefont{N.}} \&
  \bibinfo{author}{\bibnamefont{Stegemann}, \bibfnamefont{G.}}
  [\bibinfo{year}{2008}] ``\bibinfo{title}{Extended time delayed feedback
  control of stochastic dynamics in a resonant tunneling diode},''
  \emph{\bibinfo{journal}{phys.~stat.~sol. (c)}}
  \textbf{\bibinfo{volume}{5}}(\bibinfo{number}{1}), \bibinfo{pages}{194}.

\bibitem[{\citenamefont{Sch{\"o}ll \& Schuster}(2008)}]{SCH07}
\bibinfo{editor}{\bibnamefont{Sch{\"o}ll}, \bibfnamefont{E.}} \&
  \bibinfo{editor}{\bibnamefont{Schuster}, \bibfnamefont{H.~G.}} (eds.).
  [\bibinfo{year}{2008}] \emph{\bibinfo{title}{Handbook of Chaos Control}}
  (\bibinfo{publisher}{Wiley-VCH}, \bibinfo{address}{Weinheim})
  \bibinfo{note}{second completely revised and enlarged edition}.

\bibitem[{\citenamefont{Socolar} \emph{et~al.}(1994)\citenamefont{Socolar,
  Sukow, \& Gauthier}}]{SOC94}
\bibinfo{author}{\bibnamefont{Socolar}, \bibfnamefont{J.~E.~S.}},
  \bibinfo{author}{\bibnamefont{Sukow}, \bibfnamefont{D.~W.}} \&
  \bibinfo{author}{\bibnamefont{Gauthier}, \bibfnamefont{D.~J.}}
  [\bibinfo{year}{1994}] ``\bibinfo{title}{Stabilizing unstable periodic orbits
  in fast dynamical systems},'' \emph{\bibinfo{journal}{Phys.~Rev.~E}}
  \textbf{\bibinfo{volume}{50}}, \bibinfo{pages}{3245}.

\bibitem[{\citenamefont{Stegemann} \emph{et~al.}(2006)\citenamefont{Stegemann,
  Balanov, \& Sch{\"o}ll}}]{STE05a}
\bibinfo{author}{\bibnamefont{Stegemann}, \bibfnamefont{G.}},
  \bibinfo{author}{\bibnamefont{Balanov}, \bibfnamefont{A.~G.}} \&
  \bibinfo{author}{\bibnamefont{Sch{\"o}ll}, \bibfnamefont{E.}}
  [\bibinfo{year}{2006}] ``\bibinfo{title}{Delayed feedback control of
  stochastic spatiotemporal dynamics in a resonant tunneling diode},''
  \emph{\bibinfo{journal}{Phys.~Rev.~E}} \textbf{\bibinfo{volume}{73}},
  \bibinfo{pages}{016203}.

\bibitem[{\citenamefont{Unkelbach} \emph{et~al.}(2003)\citenamefont{Unkelbach,
  Amann, Just, \& Sch{\"o}ll}}]{UNK03}
\bibinfo{author}{\bibnamefont{Unkelbach}, \bibfnamefont{J.}},
  \bibinfo{author}{\bibnamefont{Amann}, \bibfnamefont{A.}},
  \bibinfo{author}{\bibnamefont{Just}, \bibfnamefont{W.}} \&
  \bibinfo{author}{\bibnamefont{Sch{\"o}ll}, \bibfnamefont{E.}}
  [\bibinfo{year}{2003}] ``\bibinfo{title}{Time--delay autosynchronization of
  the spatiotemporal dynamics in resonant tunneling diodes},''
  \emph{\bibinfo{journal}{Phys.~Rev.~E}} \textbf{\bibinfo{volume}{68}},
  \bibinfo{pages}{026204}.

\end{thebibliography}

\end{document}